\documentclass[aps,prl,twocolumn,superscriptaddress,floatfix,longbibliography,10pt]{revtex4-2}
\usepackage[utf8]{inputenc}
\DeclareUnicodeCharacter{2212}{\textminus}
\usepackage{amsmath,amssymb,ulem}
\usepackage{graphicx}
\usepackage{xcolor}
\usepackage[colorlinks=true,linkcolor=blue,citecolor=blue,urlcolor=blue]{hyperref}

\usepackage{xr-hyper}

\begin{document}

\title{Topological Hybridization Gaps from Molecular Orbitals Coupled to Dirac Fermions}

\author{Julián A. Castro}
\email{juliancastro@integra.cnea.gob.ar}
\affiliation{DFMC, GIyA, CAC, CNEA, INN (CNEA-CONICET)}
\author{Jorge I. Facio}
\email{jorge.facio@ib.edu.ar}
\affiliation{ Centro Atómico Bariloche, Instituto de Nanociencia y Nanotecnología (CNEA-CONICET) and Instituto Balseiro, Bariloche, Argentina}
\author{Verónica Vildosola}
\email{veronicavildosola@integra.cnea.gob.ar}
\affiliation{DFMC, GIyA, CAC, CNEA, INN (CNEA-CONICET)}

\date{\today}

\begin{abstract}
A magnetically gapped topological-insulator surface carries a half-quantized Hall response. 
We show that hybridizing the surface Dirac fermions with a flat band of chiral orbitals adds an integer to it: spin–momentum locking forces the hybridization to wind around the band-crossing ring, opening a topological gap. 
For orbitals with total angular momentum $j$, rotational symmetry fixes the surface Hall conductivity to $\sigma\pm j\,e^2/h$, with the sign set by the Dirac branch crossed and independent of the sign of the Dirac mass.
The result survives crystal-field quenching of the orbital moment, with the winding then set by interference between hybridization pathways. First-principles calculations for a monolayer of nitric oxide adsorbed on Bi$_2$Se$_3$ predict a $\sim20$\,meV hybridization gap and an intrinsic surface Hall conductivity of $\tfrac32\,e^2/h$, identifying molecular orbital angular momentum as a handle on surface topology.
\end{abstract}

\begin{abstract}
A magnetically gapped topological-insulator surface carries a half-quantized Hall response. 
We show that a flat band of adsorbate orbitals hybridizing with the Dirac cone of a topological-insulator surface open a gap that adds an integer to the Hall response. 
With rotational symmetry the total is Hall conductivity at the hybridization gap is $\pm j\,e^2/h$ for either sign of the Dirac mass, with $j$ the angular-momentum projection of the adsorbate orbital. When crystal fields quench the orbital moment, the two chiral components of the orbital have equal weight, yet they couple to the Dirac states with different powers of momentum. Rooted in this asymmetry, the hybridization gap remains topological. First-principles calculations for NO on Bi$_2$Se$_3$ give $\tfrac32\,e^2/h$.
\end{abstract}

\begin{abstract}
A magnetically gapped topological-insulator surface has a half-quantized Hall response. We show that a flat band of adsorbate orbitals hybridizing with its Dirac cone opens a gap that adds an integer to it. With rotational symmetry, the total is $\sigma=\pm j\,e^2/h$ for either sign of the Dirac mass, with $j$ the angular-momentum projection of the orbital. The topology is preserved when crystal fields quench the orbital moment because even if the two chiral components of the quenched orbital have equal weight, they couple asymmetrically to the different components of the Dirac states. First-principles calculations for a monolayer of nitric oxide adsorbed on Bi$_2$Se$_3$ predict $\sigma=\tfrac32\,e^2/h$.
\end{abstract}

\maketitle

\textit{Introduction.}---
Breaking time-reversal symmetry at the surface of a three-dimensional topological insulator (TI) can gap its Dirac cone, producing a half-integer-quantized anomalous Hall conductivity~\cite{haldane1988,RevModPhys.82.3045,PhysRevB.78.195424,PhysRevLett.102.146805,yu2010quantized,chang2013}. While the bulk topology fixes the half-integer part of this response, the surface electronic structure can contribute an additional integer multiple of $e^2/h$, allowing surface Hall conductivities  $\sigma_s=(n+\tfrac12)e^2/h$~\cite{PhysRevB.98.245117}.

Adsorbed molecules provide a chemically tunable route to engineer topological-insulator surface states~\cite{CO-Bi2Se3-exp,mole-TI-spin-texture,organic-mol-topo-states,molecule-TI,mole-TI-Tu2016,mole-TI-COpc_Bi2Se3,mol-TI-Valenzuela,mole-TI-2026}. Here, we propose their orbital degrees of freedom as a means of controlling the surface integer contribution to the Hall response. Specifically, the orbital character of an adsorbate can determine the topology of the bands arising from hybridization with the surface states. 
The proposal is motivated by the relation between Chern numbers and the total angular momentum of the ground state in rotationally symmetric insulators~\cite{PhysRevB.86.115112}.

We first establish how this relation operates when a band of molecular orbitals hybridizes with a surface Dirac fermion in the rotationally invariant limit. In this case, the molecular orbitals are chiral and have a well-defined angular-momentum projection $j$ along the surface normal. We show that a flat band crossing the upper Dirac branch couples to it with a hybridization that winds $j-\tfrac12$ times around the crossing ring, opening a gap with $\sigma_s=j\,e^2/h$, independent of the sign of the Dirac mass. 

In real adsorbates, however, crystal fields typically quench the orbital moment. 
We show that the chiral limit remains relevant: the opposite angular-momentum components of a quenched orbital couple asymmetrically to the Dirac states, so that one of them sets the Hall response. 
As a proof of principle, we identify an adlayer of nitric oxide (NO) molecules on Bi$_2$Se$_3$~\cite{Hsieh2009,Zhang2009,Xia2009} as a candidate for the quenched case, for which density-functional theory (DFT) predicts  $\sigma_s\simeq\tfrac32\,e^2/h$.

\textit{Basic mechanism.}---We consider a surface Dirac fermion with velocity $v$ and exchange mass $m$,
\begin{equation}
H_D=v(k_x\sigma_x+k_y\sigma_y)+m\sigma_z ,
\label{eq:Dirac}
\end{equation}
written in the basis of the Kramers doublet $\{|\psi_{+\frac12}\rangle,|\psi_{-\frac12}\rangle\}$ of the surface state at $\bar\Gamma$, with total angular momentum projection $J_z=\pm\tfrac12$ about the surface normal. Its upper and lower branches have energies $\pm E(k)$, $E=\sqrt{m^2+v^2k^2}$. Inspired by the singly occupied orbitals of open-shell molecules, we consider a spin-polarized flat band of orbitals $|\chi\rangle$, at energy $\epsilon$, with $J_z=j=m_\ell+s$, where $m_\ell$ and $s$ are the orbital and spin momentum projections. 

\begin{figure}[t]
\centering
\includegraphics[width=\columnwidth]{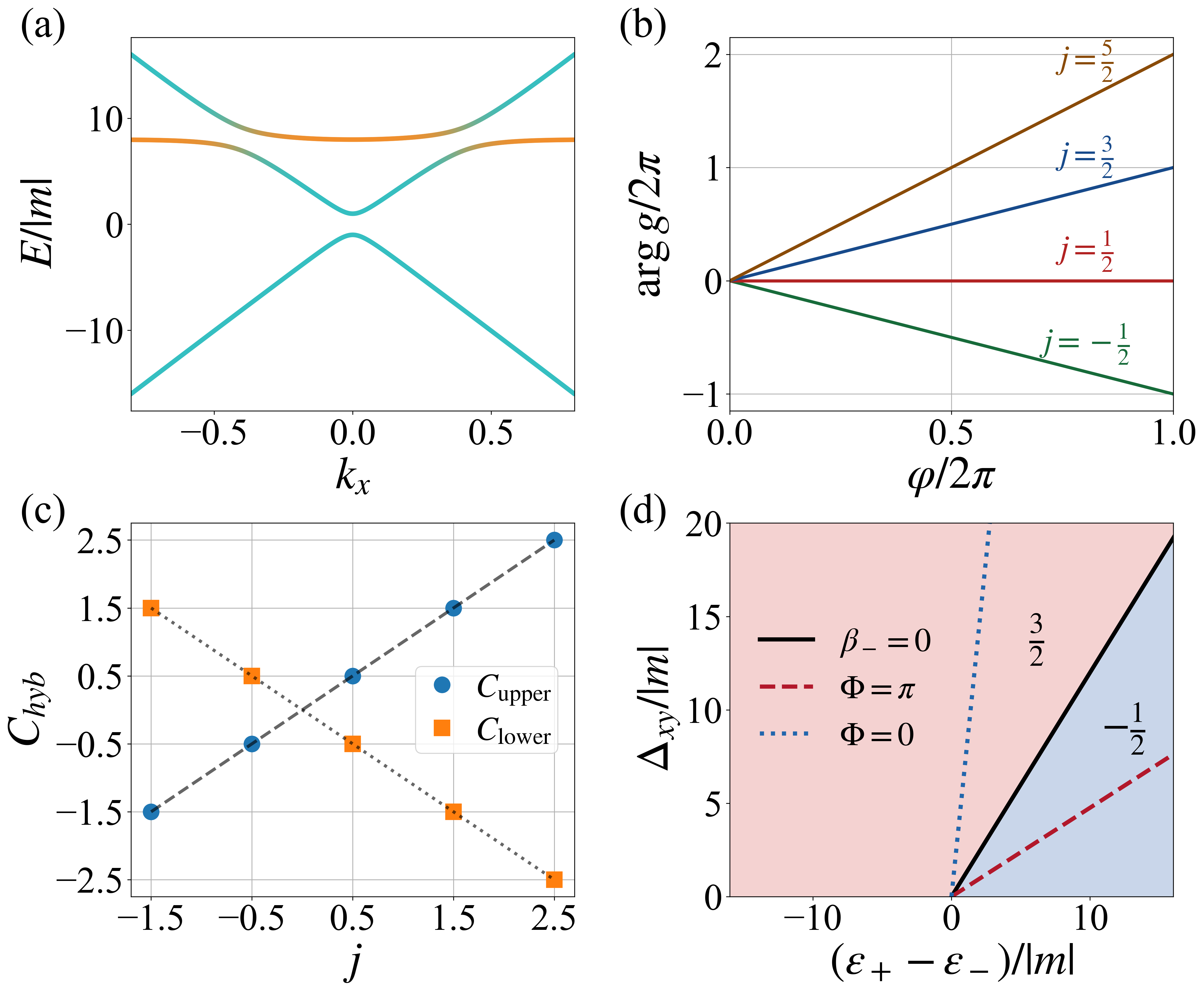}
\caption{(a) Bands of a gapped Dirac cone (teal) hybridized with a flat band of chiral $j=\tfrac32$ orbitals (orange). (b) Phase of the hybridization $g$ for different values of $j$. (c) Chern number below the hybridization gap when the chiral band crosses the upper (circles) or lower (squares) Dirac branch. (d) Phase diagram of the four-band model in the space of $\varepsilon_+-\varepsilon_-$ and $\Delta_{xy}$. In the red (blue) region, $C_{\rm hyb}=\tfrac32$ ($-\tfrac12$ ) for $\beta_\pm=0$. The dashed (dotted) lines show the boundary for a constant $\beta_-$ interfering destructively (constructively) with $\alpha_-$.}
\label{fig:mechanism}
\end{figure}

In the basis $\{|\psi_{+\frac12}\rangle,|\psi_{-\frac12}\rangle,|\chi\rangle\}$, the Bloch Hamiltonian $H(\mathbf k)$ has $H_D$ in the doublet block, $\epsilon$ on the diagonal of $|\chi\rangle$, and off-diagonal elements that couple $|\chi\rangle$ to the doublet. We take $\epsilon>|m|$ so that the chiral band crosses the upper Dirac branch on a ring $k_r$ at $E(k_r)=\epsilon$ [Fig.~\ref{fig:mechanism}(a)]. Time-reversal is broken by $m$ and by the spin polarization of the flat band; the question is the topology of the hybridization gap. 

With continuous rotational symmetry $C_\infty$ about the surface normal, $J_z$ is conserved, and a coupling matrix element between states differing by $\Delta J_z$ carries the angular factor $e^{i\Delta J_z\varphi}$, where $\mathbf k=k(\cos\varphi,\sin\varphi)$~\cite{bernevig2006quantum,SM}. The orbital therefore couples to the two components of the doublet as
\begin{equation}
\langle\psi_{\pm\frac12}|H|\chi\rangle=\tau_\pm(k)\,e^{i(j\mp\frac12)\varphi},
\label{eq:tpm-general}
\end{equation}
with $\tau_\pm\propto k^{|j\mp\frac12|}$ at small $k$. In the periodic gauge, the upper branch Dirac state is  $|u_+\rangle=\cos\tfrac\theta2|\psi_{+\frac12}\rangle+e^{i\varphi}\sin\tfrac\theta2|\psi_{-\frac12}\rangle$, with $\cos\theta=m/E$ and the hybridization with $|\chi\rangle$ is
\begin{equation}
g=\langle u_+|H|\chi\rangle=\lambda(k)\,e^{in\varphi},\qquad n=j-\tfrac12,
\label{eq:g-general}
\end{equation}
with $\lambda=\cos\tfrac\theta2\,\tau_++\sin\tfrac\theta2\,\tau_-$. The two routes from  $|\chi\rangle$ to the upper Dirac branch,  through $|\psi_{+\frac12}\rangle$ and $|\psi_{-\frac12}\rangle$, contribute to the same angular harmonic: the relative phase between the corresponding matrix elements in Eq.~\eqref{eq:tpm-general} is compensated by the factor $e^{-i\varphi}$ in $\langle u_+|$. 
Thus, for $\lambda(k_r)\neq0$, the hybridization winds $n$ times around the ring, regardless of the microscopic couplings [Fig.~\ref{fig:mechanism}(b)].

Near the ring, the lower Dirac branch is off resonance by $\simeq2\epsilon$ and can be eliminated perturbatively~\cite{SM}, giving
\begin{equation}
H_{\rm eff}=\begin{pmatrix}E(k)&g(\mathbf k)\\g^*(\mathbf k)&\epsilon\end{pmatrix}
\label{eq:Heff}
\end{equation}
in the basis $\{|u_+\rangle,|\chi\rangle\}$, where $\epsilon$ now includes a small real level shift. The lower eigenstate of $H_{\rm eff}$, which we call the intermediate band, is
$|\chi_-\rangle=-e^{i\psi}\sin\tfrac\Theta2|u_+\rangle+\cos\tfrac\Theta2|\chi\rangle$, with $\psi=\arg g=n\varphi+\arg\lambda(k)$ and $\cos\Theta=(E-\epsilon)/\sqrt{(E-\epsilon)^2+4|\lambda|^2}$. Its weight on the cone, $W=\sin^2\tfrac\Theta2$, decreases from $1$ inside the ring to $0$ outside it: a band inversion. 
Its Berry connection reads
\begin{equation}
\mathcal A_\varphi=i\langle\chi_-|\partial_\varphi\chi_-\rangle = W(k)\bigl[\mathcal A_{+}(k)-n\bigr],
\label{eq:connection}
\end{equation}
where $\mathcal A_+=-\tfrac12\Bigl(1-\frac mE\Bigr)$ is the connection of the upper Dirac branch. 
Because of the rotational symmetry, $\mathcal A_\varphi$ depends only on $k$, so the Berry curvature $\Omega=k^{-1}\partial_k\mathcal A_\varphi$ is a total derivative. Its integral reduces to boundary terms at $k=0$ and $k\to\infty$, and the Chern number of the intermediate band is
\begin{equation}
C_{\rm band}=n-\mathcal A_+(0)=j-\tfrac12\operatorname{sgn}(m).
\label{eq:Cband}
\end{equation}
Equation~\eqref{eq:Cband} is the continuum form of the relation between Chern numbers and rotation eigenvalues~\cite{PhysRevB.86.115112}. 
In a lattice with $C_n$ symmetry, the Chern number of a band is fixed, modulo $n$, by the rotation eigenvalues of its states at $\bar\Gamma$ and the other $C_n$-invariant $\mathbf k$ points on the Brillouin-zone boundary. 
In the $C_\infty$-symmetric continuum, the latter are effectively replaced by $k\to\infty$. 
The rotation eigenvalues $e^{-iJ_z\varphi}$ then determine $J_z$ itself, and the relation becomes exact~\cite{PhysRevA.98.023842}. 
Thus a band inversion that makes $J_z$ at $\bar\Gamma$ differ from its value at $k\to\infty$ yields a nonzero Chern number. 
Here, the intermediate band is molecular at large $k$, with $J_z=j$, and Dirac-like at $\bar\Gamma$, with $J_z=\pm\tfrac12$ for $m\gtrless0$, so that $C_{\rm band}=J_z(\infty)-J_z(\bar\Gamma)=j\mp\tfrac12$.

Adding the occupied lower Dirac branch, which carries the half-quantized response $C_{\rm mag}=\tfrac12\operatorname{sgn}(m)$, gives 
\begin{equation}
C_{\rm hyb}=C_{\rm mag}+C_{\rm band}=j.
\label{eq:main}
\end{equation}
Thus, the hybridization gap is topological and its Chern number depends only on the angular momentum of the chiral orbital, not on the sign of the Dirac mass or on the size of the couplings. A spin-polarized $s$ or $p_z$ orbital ($j=\tfrac12$) gives $C_{\rm hyb}=\tfrac12$; a $p_{\pm1}$ orbital gives $\tfrac32$ or $-\tfrac12$; and a $d_{\pm2}$ orbital gives $\tfrac52$ or $-\tfrac32$. 
If the chiral band crosses the lower Dirac branch, the same analysis gives $C_{\rm hyb}=-j$.
Figure~\ref{fig:mechanism}(c) confirms these values numerically.  

Orbital angular momentum can generate quantum anomalous Hall phases in $p$-band optical lattices~\cite{PhysRevLett.101.186807}, rotation symmetry can enforce multiple-winding couplings at multi-Weyl nodes~\cite{Fang2012multiWeyl}, and winding couplings produce topological bands in  polaritons~\cite{PhysRevX.5.031001}, Floquet systems~\cite{PhysRevB.90.115423}, and collective   modes~\cite{PhysRevB.104.075413,Finnigan2026PlasmonMagnon}. 
Equation~\eqref{eq:main} belongs to this family: the adsorbate angular momentum adds an integer to the Chern number of a TI surface. We next show that this survives, without symmetry protection, when crystal fields quench the orbital moment.

\textit{Quenched orbital angular momentum.}---Adsorbed orbitals generally lack a well-defined angular momentum: a crystal field mixes $m_\ell$ and favors oriented orbitals. To study this case, we consider a $p$ doublet with one spin species, as for a singly occupied $\pi^\ast$ level, in the basis $\{\lvert\psi_{+\frac12}\rangle,\lvert\psi_{-\frac12}\rangle,\lvert p_{+1}\!\uparrow\rangle,\lvert p_{-1}\!\uparrow\rangle\}$, with $J_z=+\tfrac12,-\tfrac12,+\tfrac32,-\tfrac12$. The Hamiltonian is $H_Q=H_\infty+H_\Delta$, where
\begin{equation}
H_\infty =
\begin{pmatrix}
m & v k_- & \alpha_+ k_+ & \alpha_- k_- \\
v k_+ & -m & \beta_+ k_+^{2} & \beta_- \\
\alpha_+^{*} k_- & \beta_+^{*} k_-^{2} & \varepsilon_+ & \gamma k_-^{2} \\
\alpha_-^{*} k_+ & \beta_-^{*} & \gamma^{*} k_+^{2} & \varepsilon_-
\end{pmatrix},
\label{eq:Hinf}
\end{equation}
with $k_\pm=k_x\pm ik_y$, is the most general form to lowest order in $k$ that obeys $C_z(\phi)H(\mathbf k)C_z^\dagger(\phi)=H(R_\phi\mathbf k)$, with $C_z(\phi)=\exp(-iJ_z\phi)$ and $R_\phi$ a rotation by $\phi$ in momentum space.
In this way, two states differing by $\Delta J_z$ are coupled through $k_\pm^{|\Delta J_z|}$. The amplitudes $\alpha_\pm$ couple the orbital channels $m_\ell=\pm1$ to $\lvert\psi_{+\frac12}\rangle$, and $\beta_\pm$ couple them to $\lvert\psi_{-\frac12}\rangle$. The crystal field
\begin{equation}
H_{\Delta}=-\Delta_{xy}\left(\lvert p_{+1}\rangle\langle p_{-1}\rvert+\mathrm{h.c.}\right)
\label{eq:HDelta}
\end{equation}
lowers the symmetry to $C_2$ and favors an oriented orbital, e.g., $p_x$ for $\varepsilon_+=\varepsilon_-$ and $\Delta_{xy}>0$.

For $\Delta_{xy}=0$ and $\gamma=0$, the eigenstates of the molecular block of $H_Q$ are the chiral orbitals $|p_{\pm1}\!\uparrow\rangle$, with $j=m_\ell+\tfrac12=\tfrac32$ and $-\tfrac12$.
The splitting $\varepsilon_+-\varepsilon_-$ selects which of them is the singly occupied molecular orbital (SOMO), which we assume crosses the upper branch of the Dirac cone. The basic mechanism then applies directly, and $C_{\mathrm{hyb}}=m_\ell+\tfrac12$, 
where $m_\ell$ is the orbital angular momentum projection of the SOMO.

For $\Delta_{xy}\neq0$, and neglecting $\gamma$, which mixes the levels only at order $k^2$, the SOMO is $\lvert M\rangle=u\lvert p_{+1}\!\uparrow\rangle+w\lvert p_{-1}\!\uparrow\rangle$. 
It couples to the Dirac doublet through
\begin{align}
t_{+\frac12}&=\langle\psi_{+\frac12}|H_Q|M\rangle=u\alpha_+k_++w\alpha_-k_-,\label{eq:tp}\\
t_{-\frac12}&=\langle\psi_{-\frac12}|H_Q|M\rangle=u\beta_+k_+^2+w\beta_- .
\label{eq:tm}
\end{align}
Eliminating the remote molecular level and the lower Dirac branch leads again to Eq.~\eqref{eq:Heff}, now with~\cite{SM}
\begin{equation}
g=k\cos\tfrac\theta2\left[u\tilde\alpha_+e^{i\varphi}+w\tilde\alpha_-e^{-i\varphi}\right],
\label{eq:g}
\end{equation}
where
\begin{equation}
\tilde\alpha_+=\alpha_++\beta_+k\tau,\qquad
\tilde\alpha_-=\alpha_-+\beta_-\tau/k,\qquad \tau=\tan\tfrac\theta2 .
\label{eq:al}
\end{equation}
The $m_\ell=\pm1$  components of $|M\rangle$ generate the $e^{\pm i\varphi}$ harmonics of $g$, which compete, so that the winding is
\begin{equation}
n_g=\frac{1}{2\pi}\oint d\arg g=\operatorname{sgn}\big(|u\tilde\alpha_+|-|w\tilde\alpha_-|\big),
\label{eq:ng}
\end{equation}
and the derivation of Eq.~\eqref{eq:main} carries over with $n\to n_g$: $C_{\rm hyb}=n_g+\tfrac12$~\cite{SM}. Changing $n_g$ requires $|u\tilde\alpha_+|=|w\tilde\alpha_-|$, where $g$ vanishes along a direction on the ring and the hybridization gap closes. The values of the chiral limit therefore persist when the orbital angular momentum is quenched as long as the gap stays open. 
Fig.~\ref{fig:mechanism}(d) presents the phase diagram of the model in the space of $\varepsilon_+-\varepsilon_-$, which defines the chiral SOMO when $\Delta_{xy}=0$, and $\Delta_{xy}$, which quenches the orbital angular momentum. 

Is there an intrinsic asymmetry between the two $m_\ell$ components once the orbital moment is quenched? 
There is, and it comes from the coupling to the Dirac states.
Since each unit of $\Delta J_z$ costs a power of $k$, both $|p_{\pm1}\!\uparrow\rangle$ couple to $|\psi_{+\frac12}\rangle$ linearly in $k$, through $\alpha_\pm k_\pm$, but only $|p_{-1}\!\uparrow\rangle$, which shares $J_z=-\tfrac12$ with $|\psi_{-\frac12}\rangle$, couples to it without any power of $k$, through $\beta_-$, whereas $|p_{+1}\!\uparrow\rangle$ needs $\Delta J_z=2$, giving $\beta_+k_+^2$. 
On the crossing ring, the $m_\ell=-1$ channel thus reaches the Dirac branch through two routes of comparable size, combined in $\tilde\alpha_-$ [Eq.~\eqref{eq:al}], whereas for $m_\ell=+1$ the quadratic route is typically small.
Which Dirac branch is crossed determines whether the two routes interfere constructively or destructively. On the upper branch they combine with the gauge-invariant phase $\Phi=\arg(v\beta_-\alpha_-^*)$, which involves the helicity of the Dirac cone through $v$. On the lower branch, whose $|\psi_{-\frac12}\rangle$ component has the opposite sign, they combine with $\Phi+\pi$~\cite{SM}. A finite $\beta_-$ therefore shifts the phase boundary, enlarging or shrinking the $C_{\rm hyb}=\tfrac32$ region [Fig.~\ref{fig:mechanism}(d)].

\begin{figure}[t]
\centering
\includegraphics[width=\columnwidth]{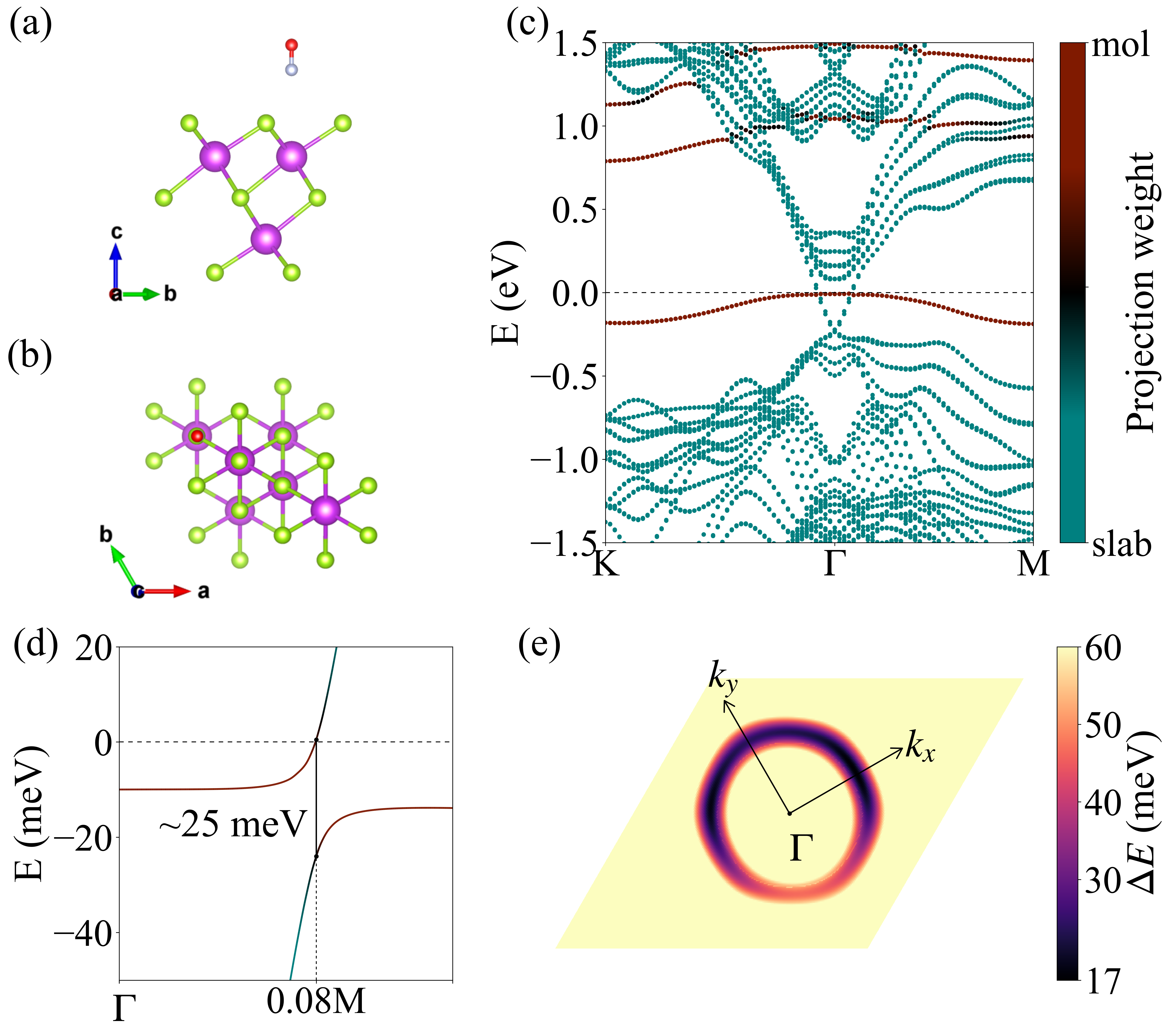}
\caption{Side (a) and top (b) view of a Bi$_2$Se$_3$ slab with one NO adsorbed overlayer. Only the top quintuple layer is shown. Green, magenta, grey and red spheres represent Se, Bi, N and O atoms, respectively. 
(c) DFT band structure projected on NO (in maroon) and on the substrate (turquoise).
(d) Wannier-interpolated band structure along $\bar\Gamma$–$\bar M$, projected onto the top surface. (e) Direct hybridization gap. }
\label{fig:struct}
\end{figure}

\begin{figure}[t]
\centering
\includegraphics[width=1\columnwidth]{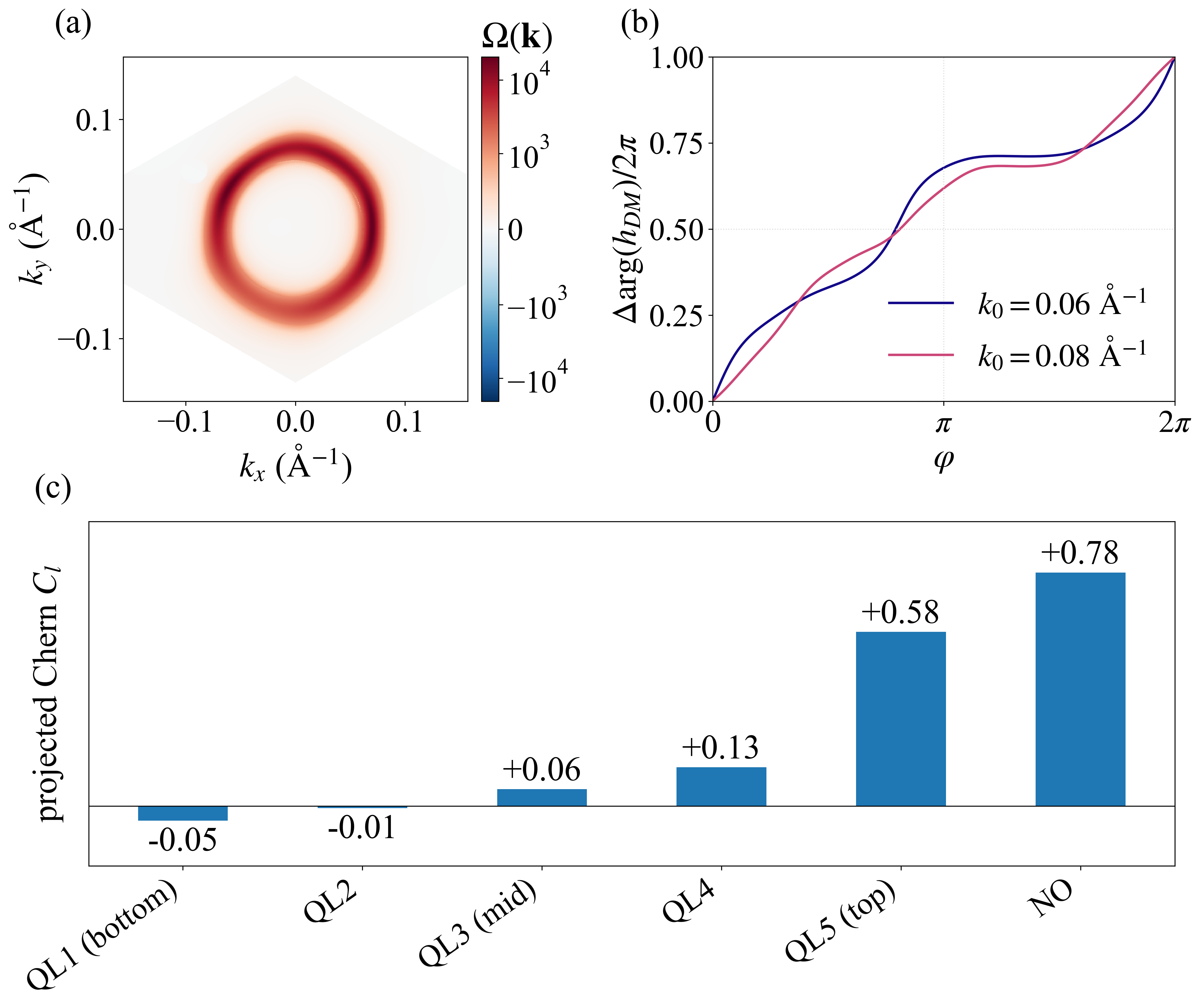}
\includegraphics[width=1\columnwidth]{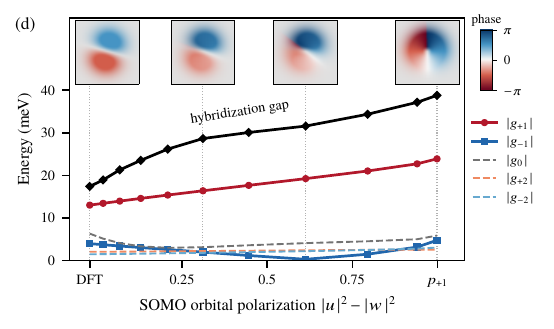}
\caption{(a) Berry curvature integrated in energy up to the hybridization gap. (b) Phase of the molecule--cone coupling $h_{DM}$ on loops of radius $k_0$ around $\bar\Gamma$. (c) Layer-resolved Chern number $C_l$. 
(d)  Hybridization gap and angular harmonics $|g_n|$ of $h_{DM}$ on the loop $k_0=0.073$~\AA$^{-1}$ as the SOMO is deformed from its DFT form to a chiral $p_{+1}$ orbital. Insets: in-plane profile of the SOMO at selected points, from a real $p_x$-like orbital with a nodal line to a $p_{+1}$ orbital whose phase winds once around the molecular axis.}
\label{fig:Berry}
\end{figure}

\textit{Candidate systems. }--- There are plenty of open-shell molecules carrying a magnetic moment which can be considered. Here we present the case of a NO adlayer on Bi$_2$Se$_3$.
NO is a radical: its antibonding $2\pi^\ast$ level is singly occupied and its frontier orbital is of $p$ type, the situation of the four-band model. 
We model an ordered NO overlayer in a uniformly spin-polarized configuration on a five-quintuple-layer Bi$_2$Se$_3$ slab, with one molecule per surface unit cell [Fig.~\ref{fig:struct}(a,b)].
As implemented in the \textit{Vienna Ab initio Simulation Package} (VASP)~\cite{VASP-1,VASP-2,VASP-3}, we use the strongly constrained and appropriately normed (SCAN) exchange and correlation functional~\cite{SCAN}, which better retains the molecular character of the orbitals than PBE~\cite{perdew1996generalized,SM}. 
In the lowest-energy configuration for the adsorption sites, the molecule stands upright atop a Bi site, with N toward the substrate.

Upon adsorption, one electron per molecule occupies a single, real $p_x$-like combination with quenched orbital moment and a spin moment perpendicular to the surface. The closest of the remaining $2\pi^\ast$ states is pushed $\approx1$\,eV higher [Fig.~\ref{fig:struct}(c)].  
The SOMO breaks $C_3$, and together with a small distortion (N displaced by 0.07\,\AA\ from the $C_3$ axis, N--O axis tilted by $2.5^\circ$), it leaves the system with no point symmetry besides the identity.

The main uncertainty of the calculation is the alignment of the SOMO with the Dirac cone, which depends on the functional. For the SCAN functional, the SOMO remains close to $\varepsilon_F$ and generates an anti-crossing on a ring with the top-surface Dirac cone, whose Dirac point lies $\sim0.2$\,eV below $\varepsilon_F$ [Fig.~\ref{fig:struct}(c,d)].
We therefore regard NO/Bi$_2$Se$_3$ as a candidate realization and use its SCAN electronic structure to demonstrate the mechanism in a realistic setting. 
The hybridization gap at the ring lies close to \(\varepsilon_F\) and is an order of magnitude larger: the direct gap varies around the ring between $\simeq17$ and $\simeq46$\,meV [Fig.~\ref{fig:struct}(d,e)]. 
Because the NO band disperses by a few meV, the indirect gap is narrower, $\simeq2$\,meV, comparable to the magnetic gap.

We construct maximally localized Wannier functions for the slab and the molecule~\cite{Wannier90}, and integrate the Berry curvature of the occupied states of the Wannier Hamiltonian $H_W(\mathbf k)$ up to different energies~\cite{SM}. 
With $\varepsilon_F$ in the magnetic gap we obtain $C_{\rm mag}\simeq-\tfrac12$, and with $\varepsilon_F$ in the hybridization gap $C_{\rm hyb}\simeq\tfrac32$; their difference is $C_{\rm band}\simeq2$ as Eq.~\eqref{eq:Cband} gives for $n=1$ and $m<0$.
The Berry curvature is concentrated around the hybridization ring [Fig.~\ref{fig:Berry}(a)] and the layer-resolved Chern number~\cite{PhysRevB.84.241106} places most of the response on the molecule and the topmost quintuple layer [Fig.~\ref{fig:Berry}(c)].
These values are not forced to be integers because the slab as a whole is metallic~\cite{gu2021spectral,mogi2022experimental,PhysRevLett.129.096601}.

To connect these results with the mechanism, we downfold the molecule--cone hybridization within the Wannier Hamiltonian $H_W(\mathbf k)$. 
We fix a reference orbital $|{\rm NO}\rangle$, the SOMO eigenvector of the isolated molecular block of $H_W(\bar\Gamma)$. 
At each $\mathbf k$ on a loop around $\bar\Gamma$, we build the projector onto the two hybridizing bands, $P(\mathbf k)$, and construct the orthonormal frame
\begin{equation}
 |M\rangle=\frac{P|{\rm NO}\rangle}{\lVert P|{\rm NO}\rangle\rVert},\qquad
 |D\rangle=\frac{(P-|M\rangle\langle M|)\,|{\rm cone}\rangle}{\lVert\cdots\rVert},
 \label{eq:frame}
\end{equation}
where $|{\rm cone}\rangle$ is the resonant cone-dominant eigenstate. 
The phase of $|D\rangle$ is fixed as in Eq.~\eqref{eq:g-general}, by taking its component along the $J_z=+\tfrac12$ state of the $\bar\Gamma$ doublet real and positive. 
The coupling $h_{DM}=\langle D|H_W|M\rangle$ is the \textit{ab initio} counterpart of $g$. 
It winds once, $n_g=+1$, on all loops around the ring [Fig.~\ref{fig:Berry}(b)], consistent with $C_{\rm hyb}=n_g+\tfrac12=\tfrac32$.

The low symmetry of NO/Bi$_2$Se$_3$ allows $h_{DM}$ to contain additional harmonics absent in $g$ of Eq.~\eqref{eq:g}. 
This raises the question of whether the $C_\infty$ limit is a meaningful reference for NO/Bi$_2$Se$_3$. To address it, we adiabatically deform the Wannier Hamiltonian so that the DFT SOMO becomes a pure $p_{+1}\!\uparrow$ orbital~\cite{SM}.
We find that both $n_g=+1$ and a finite hybridization gap are preserved along the path [Fig.~\ref{fig:Berry}(d)]. The molecule--substrate hoppings are left unchanged, so the full system does not reach $C_\infty$ symmetry. 
At the end of the path, however, the remaining harmonics $g_n$ of $h_{DM}$ satisfy $|g_{+1}|>\sum_{n\neq1}|g_n|$, so they can be removed continuously without making the coupling vanish on the ring. 
This last step acts only on the effective coupling, but it shows that the hybridization gap of NO/Bi$_2$Se$_3$ carries the winding of the $C_\infty$ limit.

The preference for $m_\ell=+1$ despite the nearly equal molecular weights on $m_\ell=\pm1$ comes from their different couplings to the Dirac states. To quantify this asymmetry, we shift the SOMO energy rigidly and fit the \textit{ab initio} harmonics to the ring-radius dependence of Eq.~\eqref{eq:al}~\cite{SM}. On the physical ring, $k_r\simeq0.074$~\AA$^{-1}$, the $m_\ell=-1$ channel has contributions $|\alpha_-|k_r\simeq22$~meV and $|\beta_-|\simeq30$~meV with relative phase $\Phi\simeq\pi$, producing substantial destructive interference. In contrast, the $\beta_+$ contribution is small compared with $\alpha_+k_r$. 
The resulting harmonics are $|g_{+1}|\simeq13$~meV and $|g_{-1}|\simeq4$~meV: the $m_\ell=-1$ channel, the only one with a constant coupling, is suppressed by destructive interference.

\textit{Conclusions.}--- 
A band of chiral orbitals with angular-momentum projection $J_z=j$ hybridizing with a surface Dirac cone opens a topological gap with Hall conductivity $+je^2/h$ for the upper branch and $-je^2/h$ for the lower one. This provides an orbital design rule, with $s$, $p$, and $d$ orbitals inducing $|C_{\rm hyb}|$ up to $\tfrac12$, $\tfrac32$, and $\tfrac52$, respectively. 
As illustrated by NO adsorption on Bi$_2$Se$_3$, the response can survive crystal-field quenching of the orbital moment because, even if the molecular orbital has equal weight on components of opposite orbital angular momentum, these couple differently to a given Dirac branch, allowing one chiral orbital to dominate the hybridization winding.

Orbital hybridization can produce gaps larger than the magnetic Dirac gap, while molecular level alignment controls the crossing energy and the Dirac branch involved. The diversity of molecular adsorbates thus offers opportunities to engineer both the magnitude and sign of the surface Hall response. 
Experimental realization will require adsorbates that form stable, ordered configurations and introduce frontier-orbital-derived bands appropriately aligned with the surface Dirac cone.

\vspace{5mm}

We thank Sofía Scozziero for fruitful discussions. This work was supported by CONICET (PIP 2023–2025, Grant No. 11220220100330CO). We acknowledge SICyT and its administrative team for access to the Clementina XXI supercomputer through project PAD-148 (2025).
Claude Opus 5.5 (Anthropic) was used to assist with post-processing of ab initio results based on the Wannier Hamiltonian, analysis of the effective model, and manuscript editing. The authors verified all results and text and take full responsibility for the content.

The data supporting this study are available in the article’s Supporting Information and in the Materials Cloud Archive (DFT input and output files, DOI: 10.24435/materialscloud:xx-xx), or from the authors upon reasonable request.

\nocite{VASP-1,VASP-2,VASP-3,SCAN,IVDW-D3,perdew1996generalized,Hsieh2009,Zhang2009,Xia2009,Menshchikova2011,GW-chulkov,GW-Aguilera,GW-Louie,Wannier90}
\bibliography{biblio}

\end{document}


\title{Supplementary Material for: Topological Hybridization Gaps from Molecular Orbitals Coupled to Dirac Fermions}
\author{Julián A. Castro}
\affiliation{DFMC, GIyA, CAC, CNEA, INN (CNEA-CONICET)}

\author{Jorge I. Facio}
\affiliation{ Centro Atómico Bariloche, Instituto de Nanociencia y Nanotecnología (CNEA-CONICET) and Instituto Balseiro, Bariloche, Argentina}

\author{Verónica Vildosola}
\affiliation{DFMC, GIyA, CAC, CNEA, INN (CNEA-CONICET)}

\maketitle
\tableofcontents

\subsection{Effective models}
\label{sec:SI-models}
This section derives the model equations of the main text. Throughout, $k_\pm=k_x\pm ik_y=ke^{\pm i\varphi}$.

\subsubsection{Selection rules from rotational symmetry [Eqs.~(2) and (8) ]}
\label{sec:SI-selection}
\label{sec:SI-eq3-7}

Under continuous rotations about the surface normal, a rotation by an angle $\phi$ is generated by $J_z$, $C_z(\phi)=\exp(-iJ_z\phi)$, so that on a state with $J_z=j$ it acts as a phase, $C_z(\phi)|j\rangle=e^{-ij\phi}|j\rangle$. The rotation also moves the crystal momentum, $\mathbf k\to R_\phi\mathbf k$, and the symmetry of the Hamiltonian reads $C_z(\phi)H(\mathbf   k)C_z^\dagger(\phi)=H(R_\phi\mathbf k)$.
 
Consider a matrix element between two states of definite angular momentum, $f(\mathbf k)=\langle j_a|H(\mathbf k)|j_b\rangle$. Using the symmetry relation and the action of $C_z$ on each state,  $\langle j_a|C_z(\phi)=e^{-ij_a\phi}\langle j_a|$ and $C_z^\dagger(\phi)|j_b\rangle=e^{ij_b\phi}|j_b\rangle$, we obtain
\begin{equation}
  f(R_\phi\mathbf k)=\langle j_a|C_z(\phi)H(\mathbf k)C_z^\dagger(\phi)|j_b\rangle=e^{i\Delta J_z\phi}f(\mathbf k),\qquad \Delta J_z=j_b-j_a .
\end{equation}
Rotating a momentum of modulus $k$ from the $k_x$ axis to the polar angle $\varphi$ then gives
\begin{equation}
  f(\mathbf k)=t(k)\,e^{i\Delta J_z\varphi},
\end{equation} 
with $t(k)=f(k,0)$. 
The phase of the coupling thus winds $\Delta J_z$ times around $\bar\Gamma$: the change of angular momentum is transferred to the angular dependence of the matrix element.
 
The angular factor also fixes how the coupling depends on $k$. In a basis of $\mathbf k$-independent orbitals, $H(\mathbf k)$ is a smooth function of $k_x$ and $k_y$, and can be expanded in powers of $k_\pm=k_x\pm ik_y=ke^{\pm i\varphi}$. 
A monomial $k_+^ak_-^b=k^{a+b}e^{i(a-b)\varphi}$ is allowed only if $a-b=\Delta J_z$, and the term of lowest order has $a+b=|\Delta J_z|$. Hence
\begin{equation}
  f(\mathbf k)\propto
  \begin{cases}
  k_+^{\Delta J_z}, & \Delta J_z>0,\\
  k_-^{|\Delta J_z|}, & \Delta J_z<0,
  \end{cases}
\end{equation}
up to corrections of relative order $k^2$, and a constant for $\Delta J_z=0$. Each unit of angular momentum exchanged costs one power of $k$. For example, the two members of the Dirac doublet differ by $\Delta J_z=-1$ and are coupled by $vk_-$, as in Eq.~(1) of the main text.

For a localized orbital $|\chi\rangle$ at energy $\epsilon$ and with $J_z=j$ and the doublet $|\psi_{\pm\frac12}\rangle$, $\Delta J_z=j\mp\tfrac12$, which gives Eq.~(2) of the main text, $\langle\psi_{\pm\frac12}|H|\chi\rangle=\tau_\pm(k)\,e^{i(j\mp\frac12)\varphi}$ with $\tau_\pm\propto k^{|j\mp\frac12|}$ at small $k$. For the four-band model, with $J_z=+\tfrac12,-\tfrac12,+\tfrac32,-\tfrac12$ for $|\psi_{+\frac12}\rangle$, $|\psi_{-\frac12}\rangle$, $|p_{+1}\uparrow\rangle$, $|p_{-1}\uparrow\rangle$, the same rule gives the entries of $H_\infty$ [Eq.~(8) of the main text]:
\begin{center}
\begin{tabular}{lcc}
\hline
element & $\Delta J_z$ & lowest order\\\hline
$\langle\psi_{+\frac12}|H|\psi_{-\frac12}\rangle$ & $-1$ & $vk_-$\\
$\langle\psi_{+\frac12}|H|p_{+1}\uparrow\rangle$ & $+1$ & $\alpha_+k_+$\\
$\langle\psi_{+\frac12}|H|p_{-1}\uparrow\rangle$ & $-1$ & $\alpha_-k_-$\\
$\langle\psi_{-\frac12}|H|p_{+1}\uparrow\rangle$ & $+2$ & $\beta_+k_+^2$\\
$\langle\psi_{-\frac12}|H|p_{-1}\uparrow\rangle$ & $0$ & $\beta_-$\\
$\langle p_{+1}\uparrow|H|p_{-1}\uparrow\rangle$ & $-2$ & $\gamma k_-^2$\\\hline
\end{tabular}
\end{center}
These rules use only $J_z$, not spin conservation. The first entry fixes the vorticity of the Dirac texture to $\eta=+1$ (Sec.~\ref{sec:model-chern}).

\subsubsection{Basic mechanism: derivation of Eqs.~(3)--(7)}
\label{sec:model-chern}

We define the Berry connection and curvature by
\begin{equation}
A_j=i\langle u|\partial_{k_j}u\rangle,\qquad
\Omega_z=\partial_{k_x}A_y-\partial_{k_y}A_x,\qquad
C=\frac{1}{2\pi}\int dk_x\,dk_y\,\Omega_z.
\label{eq:SI-conventions}
\end{equation}
To keep track of the Dirac handedness, we introduce $\eta=\pm1$ and write
\begin{equation}
H_D^{(\eta)}=v(k_x\sigma_x+\eta k_y\sigma_y)+m\sigma_z
=\begin{pmatrix}m&vke^{-i\eta\varphi}\\vke^{i\eta\varphi}&-m\end{pmatrix}.
\label{eq:SI-dirac-handedness}
\end{equation}
Here $v>0$; the Hamiltonian of Eq.~(1) of the main text has $\eta=+1$. The parameter $\eta$ is the vorticity of the pseudospin texture, i.e., its winding around $\bar\Gamma$. 
For a doublet with $J_z=\pm\tfrac12$, $C_\infty$ symmetry requires $\langle\psi_{+\frac12}|H|\psi_{-\frac12}\rangle\propto k_-$, i.e., $\eta=+1$, in a right-handed frame tied to the surface normal; $\eta=-1$ corresponds to the opposite orientation convention. We keep $\eta$ general in the Dirac part as a check on sign conventions. With $E=\sqrt{m^2+v^2k^2}$ and $\cos\theta=m/E$, we choose the eigenvectors of energy $\pm E$ of $H_D^{(+1)}$ as
\begin{equation}
|u_+\rangle=\begin{pmatrix}\cos(\theta/2)\\e^{i\varphi}\sin(\theta/2)\end{pmatrix},\qquad
|u_-\rangle=\begin{pmatrix}-e^{-i\varphi}\sin(\theta/2)\\\cos(\theta/2)\end{pmatrix}.
\label{eq:SI-dirac-spinors}
\end{equation}
They are single-valued on any circle $k>0$ (periodic gauge), and their radial connections vanish. For general $\eta$, $|u_\pm^{(\eta)}\rangle$ follow from Eq.~\eqref{eq:SI-dirac-spinors} with $\varphi\to\eta\varphi$.

Direct differentiation gives
\begin{align}
\langle u_+^{(\eta)}|\partial_\varphi u_+^{(\eta)}\rangle
 &=i\eta\sin^2(\theta/2),&
\mathcal A_{+\varphi}&=-\frac{\eta}{2}\left(1-\frac{m}{E}\right),\\
\langle u_-^{(\eta)}|\partial_\varphi u_-^{(\eta)}\rangle
 &=-i\eta\sin^2(\theta/2),&
\mathcal A_{-\varphi}&=+\frac{\eta}{2}\left(1-\frac{m}{E}\right).
\label{eq:SI-dirac-connections}
\end{align}
Since the radial connections vanish, $\Omega_\pm=k^{-1}\partial_k\mathcal A_{\pm\varphi}$, or
\begin{equation}
\Omega_\pm(k)=\mp\frac{\eta m v^2}{2(m^2+v^2k^2)^{3/2}}.
\end{equation}
For $m\ne0$, integration of the occupied lower branch gives
\begin{equation}
C_{\rm mag}=\int_0^\infty k\,dk\,\Omega_-(k)
=\frac{\eta m}{2}\int_0^\infty\frac{v^2k\,dk}{(m^2+v^2k^2)^{3/2}}
=\frac{\eta}{2}\operatorname{sgn}(m).
\label{eq:SI-magnetic-response}
\end{equation}

With the couplings of Eq.~(2) of the main text and the spinors of Eq.~\eqref{eq:SI-dirac-spinors}, the chiral orbital couples to the two branches as
\begin{align}
g&=\langle u_+|H|\chi\rangle=\cos\tfrac\theta2\,\langle\psi_{+\frac12}|H|\chi\rangle+e^{-i\varphi}\sin\tfrac\theta2\,\langle\psi_{-\frac12}|H|\chi\rangle
=\lambda(k)\,e^{i(j-\frac12)\varphi},&
\lambda&=\cos\tfrac\theta2\,\tau_++\sin\tfrac\theta2\,\tau_-,
\label{eq:SI-g-j}\\
\ell&=\langle u_-|H|\chi\rangle=-e^{i\varphi}\sin\tfrac\theta2\,\langle\psi_{+\frac12}|H|\chi\rangle+\cos\tfrac\theta2\,\langle\psi_{-\frac12}|H|\chi\rangle
=\lambda'(k)\,e^{i(j+\frac12)\varphi},&
\lambda'&=-\sin\tfrac\theta2\,\tau_++\cos\tfrac\theta2\,\tau_- .
\label{eq:SI-ell-j}
\end{align}
The spinor phase $e^{\mp i\varphi}$ brings both routes into the same angular harmonic, so that $g$ winds $n=j-\tfrac12$ times and $\ell$ winds $j+\tfrac12$ times; the amplitudes $\lambda$, $\lambda'$ depend only on $k$.

Although $|\ell|/|g|$ need not be small, the lower branch is separated in energy from the hybridization gap. For $\epsilon>|m|$, the unperturbed ring satisfies $E(k_r)=\epsilon$, and that separation is $E+\epsilon=2\epsilon$. 

In the basis $\{|u_+\rangle,|u_-\rangle,|\chi\rangle\}$, $|u_-\rangle$ couples only to $|\chi\rangle$, through $\ell$. A canonical transformation, with generator (to leading order) $S'=\ell\,|u_-\rangle\langle\chi|/(E+\epsilon)-{\rm h.c.}$, removes this coupling to first order and gives the resonant block
\begin{equation}
H_{\rm eff}=\begin{pmatrix}E&g\\g^*&\widetilde\epsilon\end{pmatrix},\qquad
\widetilde\epsilon=\epsilon+\frac{|\ell|^2}{E+\epsilon}+O\!\left(\frac{|\ell|^4}{(E+\epsilon)^3}\right).
\label{eq:SI-Heff-ring}
\end{equation}
This is the $H_{\rm eff}$ of Eq.~(4) of the main text. The level shift $|\ell|^2/(E+\epsilon)$ is real and small; replacing it by a constant closes no gap, and we absorb it into $\epsilon$ (dropping the tilde).

We now derive the Berry connection of the intermediate band, i.e., the lower eigenstate of $H_{\rm eff}$, written in the basis $\{|u_+^{(\eta)}\rangle,|\chi\rangle\}$. We write
\begin{equation}
g=|g|e^{i\psi},\qquad
D=\sqrt{(E-\epsilon)^2+4|g|^2},\qquad
\cos\Theta=\frac{E-\epsilon}{D}.
\end{equation}
The eigenvalues are $(E+\epsilon\pm D)/2$. A normalized eigenstate of the lower eigenvalue is
\begin{equation}
|\chi_-\rangle=-e^{i\psi}\sin(\Theta/2)|u_+^{(\eta)}\rangle
+\cos(\Theta/2)|\chi\rangle.
\label{eq:SI-hybridized-state}
\end{equation}
Since $|\chi\rangle$ is independent of momentum and orthogonal to the cone states and their derivatives, differentiating Eq.~\eqref{eq:SI-hybridized-state} gives
\begin{align}
\langle\chi_-|\partial_q\chi_-\rangle
 &=iW\partial_q\psi+W\langle u_+^{(\eta)}|\partial_q u_+^{(\eta)}\rangle,\\
\mathcal A_q^{\chi_-}&=W\bigl(\mathcal A_{+q}-\partial_q\psi\bigr),
\label{eq:SI-hybridized-connection}
\end{align}
where $q=k$ or $\varphi$, and $W$ is the cone weight of the intermediate band,
\begin{equation}
W=\sin^2\frac{\Theta}{2}=\frac12\left[1-\frac{E-\epsilon}{\sqrt{(E-\epsilon)^2+4|g|^2}}\right].
\label{eq:SI-W}
\end{equation}
For the chiral orbital, $|g|=\lambda(k)$ and $\psi=n\varphi+\arg\lambda(k)$ [Eq.~\eqref{eq:SI-g-j}]. Hence $W=W(k)$ depends on momentum only through the cone dispersion $E(k)$ and the coupling strength $\lambda(k)$, $\partial_\varphi\psi=n$, and
\begin{equation}
\mathcal A_\varphi^{\chi_-}(k)=W(k)\bigl(\mathcal A_{+\varphi}(k)-n\bigr),
\label{eq:SI-connection-k}
\end{equation}
which is Eq.~(5) of the main text for $\eta=+1$. The radial connection, $\mathcal A_k^{\chi_-}=-W\,\partial_k\arg\lambda$, depends only on $k$ and does not contribute to the curvature, $\Omega_{\chi_-}=k^{-1}\,d\mathcal A_\varphi^{\chi_-}/dk$.

Since the curvature is a total derivative, the Chern number of the intermediate band [Eq.~\eqref{eq:SI-conventions}] reduces to boundary terms,
\begin{equation}
C_{\rm band}=\frac{1}{2\pi}\int d^2k\,\Omega_{\chi_-}
=\int_0^\infty k\,dk\,\Omega_{\chi_-}(k)
=W(\infty)\bigl(\mathcal A_{+\varphi}(\infty)-n\bigr)-W(0)\bigl(\mathcal A_{+\varphi}(0)-n\bigr).
\label{eq:SI-Cint-boundary}
\end{equation}
$C_{\rm band}$ is thus fixed by the cone weight at the two ends:
\begin{itemize}
\item At $k=0$, $\lambda(0)=0$ unless the orbital has the same $J_z$ as the band-edge state of the upper branch, $j=\tfrac12$ with $m>0$ or $j=-\tfrac12$ with $m<0$, where $\tau_+$ or $\tau_-$ is constant [Eq.~(2) of the main text]. Otherwise $W(0)=1$ if $E(0)=|m|<\epsilon$ and $W(0)=0$ if $|m|>\epsilon$. In the first case the upper Dirac branch crosses the chiral level on a ring, $E(k_r)=\epsilon$ (band inversion), and the intermediate band changes from cone-like inside the ring to molecular outside it, over a width $\sim\lambda/v$ around $k_r$. In the two exceptional cases $W(0)$ is not quantized, but $\mathcal A_{+\varphi}(0)-n=0$, so the inner boundary term vanishes, $C_{\rm band}=0$, and $C_{\rm hyb}=C_{\rm mag}=j$: the orbital merely repels the band-edge state of equal $J_z$.
\item At large $k$, the cone and the chiral level are detuned, and $g$ mixes them only weakly, by $\sim\lambda/|E-\epsilon|$. Switching $g$ off there closes no gap, and gives $W(\infty)=0$.\footnote{With $g\propto k$ at all momenta, $\lambda/(E-\epsilon)$ tends to a constant and so does $W$. In the numerical check of Fig.~1(c) of the main text we multiply the couplings by $e^{-k^2/k_c^2}$, which gives $W(\infty)=0$ without closing any gap, and compute $C$ with the Fukui--Hatsugai--Suzuki method over a square $|k_{x,y}|\le K$ with $K\gg k_c,k_r$ without identifying opposite edges;  the result reproduces $\pm j$ to within $5\times10^{-3}$ for $-\tfrac32\le j\le\tfrac52$ and both signs of $m$. As usual for a single Dirac surface, the half-integer values refer to the surface contribution; the quantity fixed by the mechanism is the change across the hybridization, $C_{\rm hyb}-C_{\rm mag}$, which is an integer.}
\end{itemize}
Without band inversion ($\epsilon<|m|$) both boundary terms vanish and $C_{\rm band}=0$. With band inversion ($\epsilon>|m|$),
\begin{equation}
C_{\rm band}=n-\mathcal A_{+\varphi}(0)=n+\frac{\eta}{2}\left[1-\operatorname{sgn}(m)\right].
\label{eq:SI-intermediate-chern}
\end{equation}
Adding the occupied lower Dirac branch gives
\begin{equation}
C_{\rm hyb}=C_{\rm mag}+C_{\rm band}
=\frac{\eta}{2}\operatorname{sgn}(m)
+n+\frac{\eta}{2}\left[1-\operatorname{sgn}(m)\right]
=n+\frac{\eta}{2}.
\label{eq:SI-hybridization-response}
\end{equation}
For $\eta=+1$ and $n=j-\tfrac12$, this is $C_{\rm hyb}=j$, Eq.~(7) of the main text. For $j=\tfrac32$ ($n=1$), the intermediate band has $C_{\rm band}=1$ when $m>0$ and $C_{\rm band}=2$ when $m<0$, while the total below the hybridization gap remains $\tfrac32$. At $m=0$ the magnetic gap closes and the separate band Chern numbers cease to be defined, but their sum below an open hybridization gap remains well defined. The result extends beyond the regime $|\ell|\ll2\epsilon$ along any deformation that preserves the hybridization gap and the separation from remote bands.

If the chiral band lies below the Dirac point, $\epsilon<-|m|$, it crosses the lower branch on the ring $-E(k_r)=\epsilon$. The coupling to that branch, $\ell$ of Eq.~\eqref{eq:SI-ell-j}, winds $n'=j+\tfrac12$ times. Eliminating the upper branch, off resonance by $\simeq2|\epsilon|$, leaves $H'_{\rm eff}=\begin{pmatrix}-E&\ell\\\ell^*&\epsilon\end{pmatrix}$ in the basis $\{|u_-\rangle,|\chi\rangle\}$. Its lower eigenstate is now molecular inside the ring and cone-like outside it, so its cone weight $W'$ goes from $W'(0)=0$ to $W'(\infty)=1$, and its connection is $\mathcal A_\varphi=W'(\mathcal A_{-\varphi}-n')$, with $\mathcal A_{-\varphi}=\tfrac12(1-m/E)$ [Eq.~\eqref{eq:SI-dirac-connections}]. The boundary terms of Eq.~\eqref{eq:SI-Cint-boundary} now come from the outer circle, and give $C=\mathcal A_{-\varphi}(\infty)-n'=\tfrac12-n'=-j$. This band is the only one below the hybridization gap, so the Hall response there is $-j$, for either sign of $m$.

\subsubsection{Quenched orbital angular momentum: derivation of Eqs.~(10)--(14)}
\label{sec:SI-eq8}
\label{sec:model-reduction}
We reduce $H_Q=H_\infty+H_\Delta$ to the effective two-level problem of Eq.~(4) by two canonical transformations. The first removes the remote molecular level; the second removes the lower Dirac branch, as in Sec.~\ref{sec:model-chern}.

We define $H_{\rm mol}$ as the molecular block of $H=H_\infty+H_\Delta$,
\begin{equation}
H_{\rm mol}=\begin{pmatrix}\varepsilon_+&\gamma k_-^2-\Delta_{xy}\\\gamma^* k_{+}^2-\Delta_{xy}^*&\varepsilon_-\end{pmatrix}.
\end{equation}
As in the main text, we set $\gamma=0$, so that the molecular eigenstates do not depend on momentum. The eigenvalues are $(\varepsilon_++\varepsilon_-)/2\pm\sqrt{(\varepsilon_+-\varepsilon_-)^2/4+|\Delta_{xy}|^2}$, denoted as $\epsilon_M$ for the lowest value and $\epsilon_R$ for the largest. The relevant SOMO in this model is $|M\rangle=u|p_{+1}\uparrow\rangle+w|p_{-1}\uparrow\rangle$, of energy $\epsilon_M$, and the other eigenstate is $|R\rangle=-w^*|p_{+1}\uparrow\rangle+u^*|p_{-1}\uparrow\rangle$, of energy $\epsilon_R$. Here $|u|^2+|w|^2=1$, and $w$ is distinct from the Dirac velocity $v$. The SOMO couples to both components $|\psi_{\pm\frac12}\rangle$ of the Dirac doublet [Eqs.~(10) and (11) of the main text],
\begin{equation}
t_{+\frac12}(\mathbf k)=\langle\psi_{+\frac12}|H_Q|M\rangle=u\alpha_+k_++w\alpha_-k_-,\qquad
t_{-\frac12}(\mathbf k)=\langle\psi_{-\frac12}|H_Q|M\rangle=u\beta_+k_+^2+w\beta_- .
\label{eq:SI-tup-tdn}
\end{equation}
Each orbital channel is therefore reached by two routes: $|p_{+1}\uparrow\rangle$ by $\alpha_+$ ($\Delta J_z=+1$) and $\beta_+$ ($\Delta J_z=+2$), and $|p_{-1}\uparrow\rangle$ by $\alpha_-$ ($\Delta J_z=-1$) and $\beta_-$ ($\Delta J_z=0$).

For $\gamma=0$, $\langle M|H_Q|R\rangle=0$, so the remote level $|R\rangle$ couples only to the Dirac doublet, and it can be removed by a Schrieffer--Wolff transformation. To leading order in $V_R/\Delta_R$, where $V_R$ is the size of the coupling to $|R\rangle$ and $\Delta_R=\epsilon_R-\epsilon_M$ its separation from the resonance, the transformation generates a correction of order $|V_R|^2/\Delta_R$ that acts only within the Dirac block, where it adds the symmetry-allowed terms of a Dirac Hamiltonian (a scalar shift, corrections to $m$ and $v$, and higher-order terms in $k$), and leaves the molecular row untouched. These corrections cannot change $C_{\rm hyb}$ as long as they do not close the hybridization gap; since $C_{\rm hyb}=j$ for either sign of $m$, this holds even if they exceed the magnetic gap. To leading order, eliminating $|R\rangle$ thus amounts to keeping the $\{\psi_{+\frac12},\psi_{-\frac12},M\}$ block of $H$:
\begin{equation}
H_3=\begin{pmatrix}m&vk_-&t_{+\frac12}\\vk_+&-m&t_{-\frac12}\\t_{+\frac12}^*&t_{-\frac12}^*&\epsilon_M\end{pmatrix}.
\label{eq:SI-H3}
\end{equation}

With the spinors of Eq.~\eqref{eq:SI-dirac-spinors},
\begin{equation}
g=\langle u_+|H_3|M\rangle=\cos(\theta/2)\,t_{+\frac12}+e^{-i\varphi}\sin(\theta/2)\,t_{-\frac12},\qquad
\ell=\langle u_-|H_3|M\rangle=-e^{i\varphi}\sin(\theta/2)\,t_{+\frac12}+\cos(\theta/2)\,t_{-\frac12} .
\end{equation}
Inserting Eq.~\eqref{eq:SI-tup-tdn} gives Eqs.~(12) and (13) of the main text,
\begin{equation}
g=k\cos(\theta/2)\left[u\tilde\alpha_+e^{i\varphi}+w\tilde\alpha_-e^{-i\varphi}\right],\qquad
\tilde\alpha_+=\alpha_++\beta_+k\tau,\qquad
\tilde\alpha_-=\alpha_-+\frac{\beta_-\tau}{k},\qquad \tau=\tan\frac\theta2=\frac{vk}{E+m}.
\label{eq:SI-g-general}
\end{equation}
Eliminating the lower branch as in Sec.~\ref{sec:model-chern} then gives $H_{\rm eff}$ of Eq.~(4) in the basis $\{|u_+\rangle,|M\rangle\}$, with $\epsilon=\epsilon_M$ up to the real level shift $|\ell|^2/(E+\epsilon_M)$.

Unlike Eq.~\eqref{eq:SI-g-j}, $g$ now contains two harmonics, and its modulus depends on $\varphi$ through their interference. On the ring, $g$ winds $n_g=\operatorname{sgn}(|u\tilde\alpha_+(k_r)|-|w\tilde\alpha_-(k_r)|)$ times, Eq.~(14) of the main text. Reducing the smaller of $|u\tilde\alpha_+|$ and $|w\tilde\alpha_-|$ continuously to zero keeps $g\neq0$ on the ring, hence keeps the gap open, and deforms $g$ to the single-harmonic form $\lambda(k)e^{in_g\varphi}$ of Eq.~\eqref{eq:SI-g-j}. Sec.~\ref{sec:model-chern} then applies with $n\to n_g$ and gives $C_{\rm hyb}=n_g+\tfrac12$. Changing $n_g$ requires $|u\tilde\alpha_+|=|w\tilde\alpha_-|$, where $g$ vanishes along a direction on the ring and the hybridization gap closes.

Finally, we comment on how the couplings depend on the choice of phases of the basis states. While the phases of $\alpha_\pm$ and $\beta_\pm$ depend on this choice, we will show that the combinations  $\Phi_\pm=\arg(v\beta_\pm\alpha_\pm^*)$ do not, and they determine whether the $\beta_\pm$ terms interfere constructively or destructively with the $\alpha_\pm$ terms. 

Since $|M\rangle=u|p_{+1}\!\uparrow\rangle+w|p_{-1}\!\uparrow\rangle$, comparing the coefficients of $u$ and $w$ in Eq.~\eqref{eq:SI-tup-tdn} identifies each coupling with a single matrix element,
\begin{equation}
    \langle\psi_{+\frac12}|H_Q|p_{\pm1}\!\uparrow\rangle=\alpha_\pm k_\pm,\qquad
    \langle\psi_{-\frac12}|H_Q|p_{+1}\!\uparrow\rangle=\beta_+k_+^2,\qquad
    \langle\psi_{-\frac12}|H_Q|p_{-1}\!\uparrow\rangle=\beta_-,
\end{equation}
and the Dirac term gives $\langle\psi_{+\frac12}|H_Q|\psi_{-\frac12}\rangle=vk_-$. 
Under the rephasing $|\psi_{+\frac12}\rangle\to e^{ia}|\psi_{+\frac12}\rangle$, $|\psi_{-\frac12}\rangle\to   e^{ib}|\psi_{-\frac12}\rangle$, $|p_{\pm1}\!\uparrow\rangle\to e^{ic_\pm}|p_{\pm1}\!\uparrow\rangle$, each matrix element acquires the phase of its ket times the conjugate phase of its bra, while $H_Q$ and $k_\pm$ are unchanged. 
Hence
\begin{equation}
    \alpha_\pm\to e^{i(c_\pm-a)}\alpha_\pm,\qquad
    \beta_\pm\to e^{i(c_\pm-b)}\beta_\pm,\qquad
    v\to e^{i(b-a)}v,
\end{equation}
while $u\to e^{-ic_+}u$ and $w\to e^{-ic_-}w$ leave $|M\rangle$ unchanged, and $t_{\pm\frac12}$ change only by the phase of their bra. The phases cancel in $v\beta_\pm\alpha_\pm^*$, so $\Phi_\pm$ is 
gauge invariant. In the gauge of Eq.~\eqref{eq:SI-dirac-spinors}, $v>0$ and $\Phi_\pm=\arg(\beta_\pm/\alpha_\pm)$ is the relative phase of the two terms of $\tilde\alpha_\pm$ in Eq.~\eqref{eq:SI-g-general}:
\begin{equation}
    |\tilde\alpha_+|=\bigl||\alpha_+|+|\beta_+|k\tau\,e^{i\Phi_+}\bigr|,\qquad
    |\tilde\alpha_-|=\Bigl||\alpha_-|+\frac{|\beta_-|\tau}{k}\,e^{i\Phi_-}\Bigr|,
\end{equation}
so $\Phi_\pm=\pi$ ($0$) gives destructive (constructive) interference.

For a SOMO crossing the lower branch, $\ell=\langle u_-|H_3|M\rangle=-k\sin\tfrac\theta2\bigl[u\hat\alpha_+e^{2i\varphi}+w\hat\alpha_-\bigr]$, with $\hat\alpha_+=\alpha_+-\beta_+k/\tau$ and $\hat\alpha_-=\alpha_--\beta_-/(k\tau)$. The $\beta_\pm$ terms now enter with the opposite sign, because the $|\psi_{-\frac12}\rangle$ component of $|u_-\rangle$ has the opposite sign relative to $|\psi_{+\frac12}\rangle$, so the relative phase is $\Phi_\pm+\pi$.

\subsection{\textit{Ab initio} calculations}
\label{sec:SI-abinitio}

\subsubsection{Computational details}
We performed first-principles calculations within density functional theory (DFT) \cite{} using the projector augmented-wave (PAW) formalism as implemented in the Vienna \textit{ab initio} Simulation Package (VASP) \cite{VASP-1,VASP-2,VASP-3}. The PAW potentials include 15, 6, 6, and 5 valence electrons for Bi, Se, O, and N, respectively. All calculations presented in the main text were performed using the strongly constrained and appropriately normed (SCAN) semilocal exchange-correlation functional\cite{SCAN}. All calculations were done including spin-orbit coupling as implemented in the VASP code\cite{SOC-in-VASP}.

\begin{figure}[h]
\centering
\includegraphics[width=0.8\columnwidth]{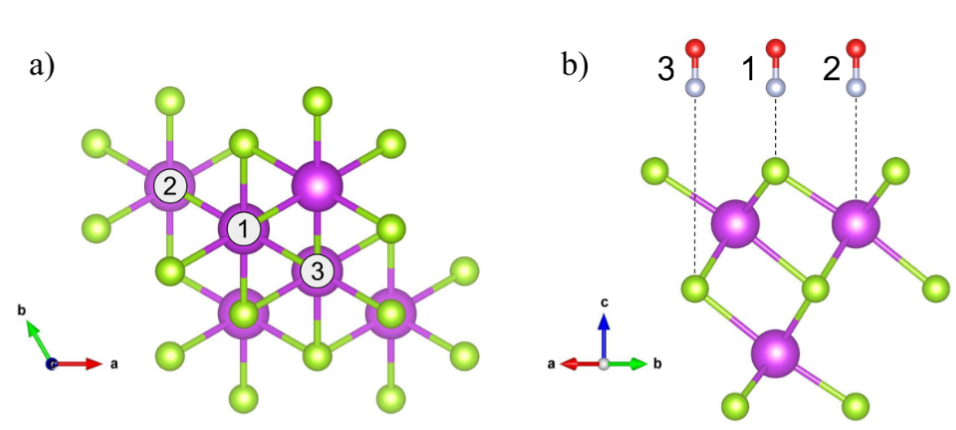}  
\caption{Adsorption sites explored.}
  \label{fig:S1}
\end{figure}

The optimized bulk lattice parameters of Bi$_2$Se$_3$ crystal are $a$ = 4.15\,\text{\AA} and $c$ = 28.7\,\text{\AA}. The bulk quintuple layers (QLs) are separated by a van der Waals gap of 2.56\,\text{\AA}. The Bi$_2$Se$_3$ (0001) surface was modeled using a five-QL slab separated from its periodic images by 15\,\text{\AA}  of vacuum. This slab thickness was verified to be sufficient to suppress hybridization between electronic states localized at the two opposite surfaces. 
A $1\times1$ surface unit cell was employed, with one NO molecule adsorbed per unit cell, corresponding to the full-coverage limit considered in the present work.
Previous first-principles studies of gas adsorption on Se-deficient Bi$_2$Se$_3$ identified NO as an ineffective vacancy passivator, in contrast to NO$_2$ and O$_2$~\cite{PhysRevLett.110.016403}. Here, we investigate NO adsorption on the pristine surface, focusing on the hybridization between its frontier molecular orbital and the topological surface states. This provides a description of the intrinsic molecule–surface coupling, relevant to restored Dirac states and to defect-free regions of realistic samples.

During structural optimization of the slab, the three uppermost QLs were fully relaxed together with the adsorbed molecule, while the two bottom QLs were kept fixed at bulk positions. The structure was relaxed until the residual forces on all relaxed atoms were smaller than 0.01\,eV/Å.

Dipole corrections were applied to remove spurious electrostatic interactions between periodically repeated slabs arising from asymmetric surface geometry and molecular adsorption. Van der Waals dispersion interactions were added to the total energy using Grimme's D3 correction with Becke–Johnson damping\cite{IVDW-D3}, providing a more accurate description of the molecule-surface interaction.

Brillouin-zone integrations were performed using a $\Gamma$-centered $13\times13\times1$ $k$-point mesh, which was found to provide converged total energies and, in particular, a stable determination of the Fermi level. A plane-wave cutoff of 450~eV was used.

Several adsorption sites and molecular orientations were considered. Figure \ref{fig:S1} shows a top view of the three adsorption sites investigated. For each site, both N-down and O-down molecular orientations were considered as initial configurations. Among the configurations examined, the most energetically favorable structure corresponds to adsorption at site 2, on top of the Bi site, with the N atom pointing towards the surface in an upward position. 
 Details of binding energies and structural configurations of all the studied cases will be published elsewhere. 

\begin{figure}[t]
\centering
\includegraphics[width=0.95\columnwidth]{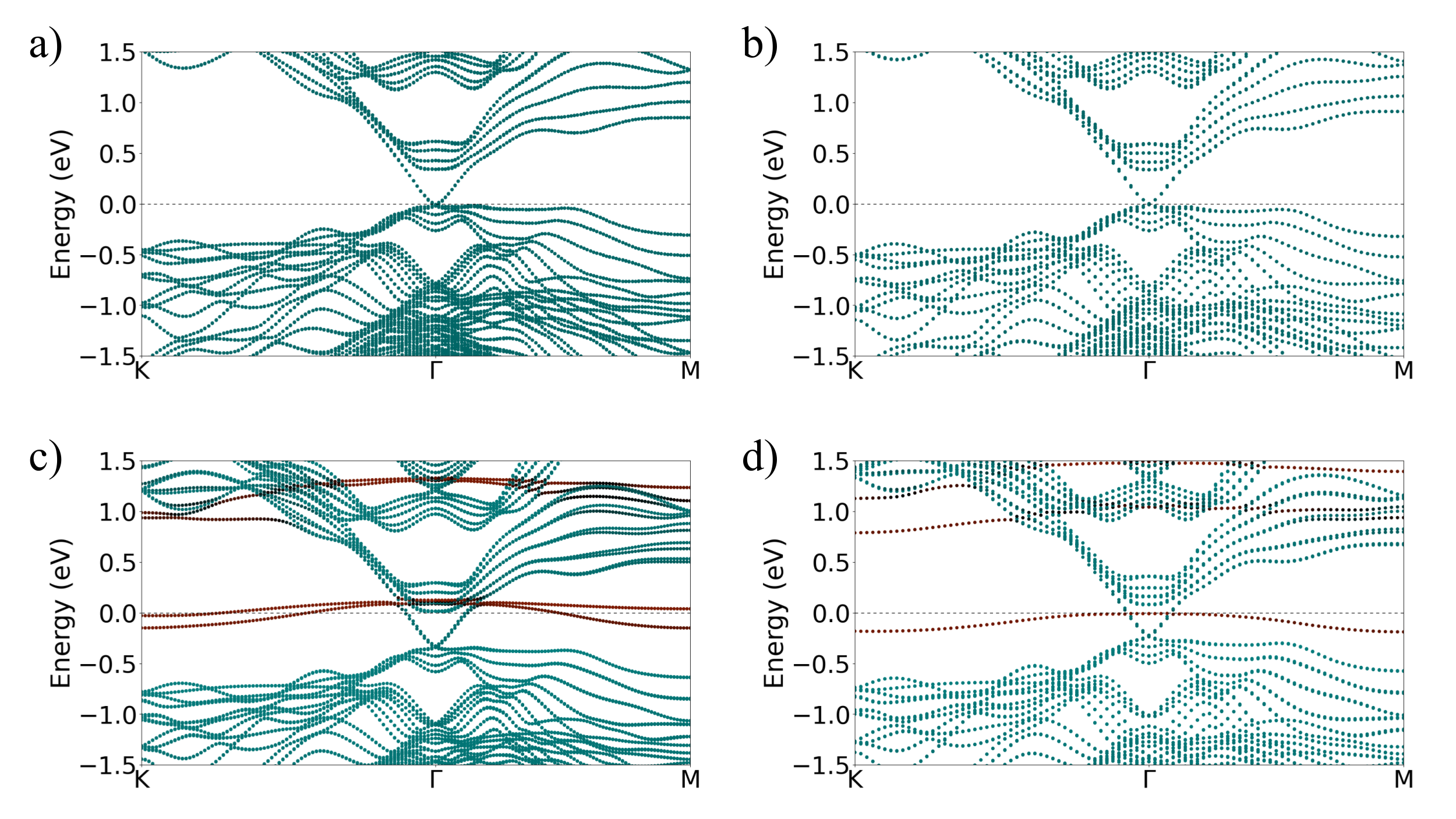}  
\caption{a) PBE b) SCAN bandstructure for a 5QL slab of Bi$_2$Se$_3$ (0001), c) PBE and d) SCAN bandstructure obtained after NO adsorption. Brown (turquoise) bands indicate molecule(substrate) projections.}
  \label{fig:S2}
\end{figure}

Standard semilocal functionals such as PBE \cite{perdew1996generalized} describe the electronic structure and topological surface states of Bi$_2$Se$_3$ reasonably well \cite{Hsieh2009,Zhang2009,Xia2009,Menshchikova2011}. However, they can overestimate the binding of molecular adsorbates and therefore may lead to an inaccurate description of the molecule–surface interaction. 
This issue is particularly relevant for the open-shell NO molecule. For instance, the calculated value of the SOMO-LUMO gap of the isolated molecule obtained within PBE is 0.2\,eV, making it comparable to the Bi$_2$Se$_3$ bulk gap and intermolecular hopping.
PBE and SCAN give very similar band structures for the pristine Bi$_2$Se$_3$ slab [Fig.~\ref{fig:S2}(a,b)]. Upon NO adsorption, the small molecular gap obtained within PBE results in a pronounced broadening and hybridization of the molecular-derived bands with the electronic states of the Bi$_2$Se$_3$ substrate where both SOMO and LUMO orbitals participate in bonding after adsorption.  
Within the SCAN functional, the SOMO-LUMO gap of the isolated molecule turns out to be 1.1\,eV—five times larger than that obtained with PBE. With SCAN, the molecular-derived bands remain much narrower and retain their molecular character, suppressing artificial overhybridization and leading to a more reliable description [Fig.~\ref{fig:S2}(c,d)].

Last, many-body effects are known to introduce important changes in the bulk band-gap structure of Bi$_2$Se$_3$ \cite{GW-chulkov,GW-Aguilera,GW-Louie}. However, we expect them to have a comparatively minor effect on the hybridization between the dispersive edge states and the molecular orbital. A detailed assessment of these effects is beyond the scope of the present work.

\begin{figure}[h!]
\centering
\includegraphics[width=0.9\columnwidth]{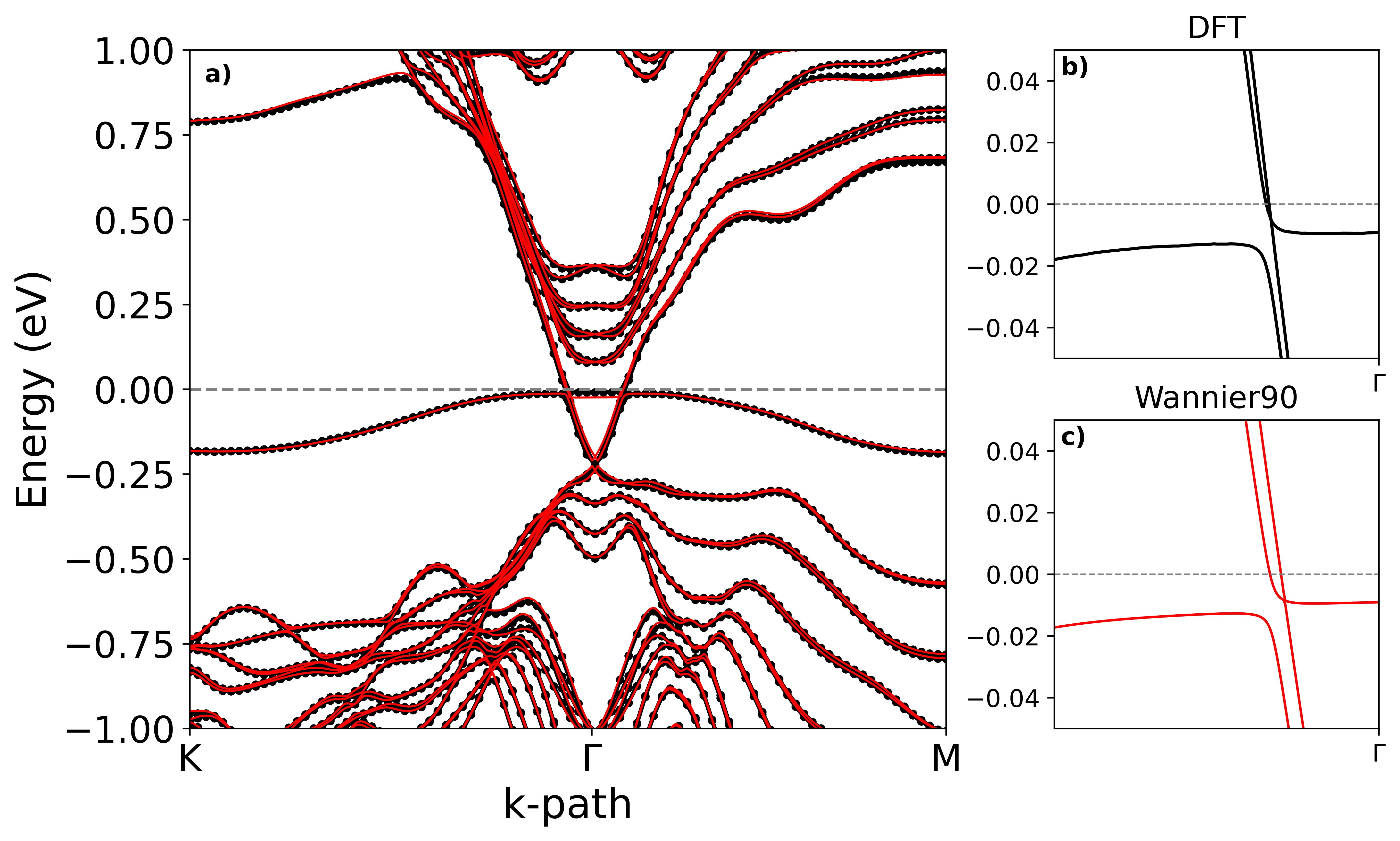}  
\caption{a) Kohn-Sham (SCAN) (black) and Wannier interpolated (red) bandstructures obtained for NO adsorption on Bi$_2$Se$_3$. (b) zoom of the Kohn-Sham (SCAN) bandstructure within the hybridization energy region. (c) Same as (b) for the interpolated Wannier Hamiltonian. }
\label{fig:PBE-Wannier-bands}
\end{figure}

In Fig. \ref{fig:PBE-Wannier-bands}(a), the interpolated Wannier bandstructure (red curve) was constructed using VASP interfaced with Wannier90~\cite{Wannier90}, including \(p\)-orbitals of all atoms in the system (slab+molecule) as the initial projection set, resulting in a total of 162 Wannier functions \cite{Wannier90}. A frozen (inner) energy window of \(E_{\mathrm{froz}}\in[-1.1,+4.5]\)\,eV with the Fermi energy at 2.4 eV was employed to ensure an accurate interpolation of the relevant electronic states.  In Fig.~\ref{fig:PBE-Wannier-bands}(b) a zoom of the hybridization energy region of the Kohn-Sham (SCAN) bandstructure is shown. The non hybridized Dirac cone can also be observed in this panel. It was projected out for clarity in Fig.~2(d) of the main text. In the panel (c) of Figure \ref{fig:PBE-Wannier-bands} we show the corresponding zoom of the Wannier bands.  

\subsubsection{Berry Curvature from the Wannier Hamiltonian}
\label{sec:wannier-berry-curvature}
The Berry curvature was calculated using the Wannier-interpolated Hamiltonian obtained from the first-principles calculations described above. 
The calculations were performed using Wannier90, evaluating the Berry curvature with the Kubo formula on a two-dimensional \(500\times500\) uniform \(k\)-point mesh over a reduced region of the surface Brillouin zone, defined by \(\mathbf{k}_{\mathrm{corner}}=(-0.08,-0.08,0)\,\text{\AA}^{-1}\), \(\mathbf{b}_1=(0.16,0,0)\,\text{\AA}^{-1}\), and \(\mathbf{b}_2=(0,0.16,0)\,\text{\AA}^{-1}\). The integration region was chosen to encompass the region where the Berry curvature is strongly localized around the hybridization gap. In particular, more than 99\% of the total Berry curvature contribution is contained within this region, as the Berry curvature decays rapidly away from the hybridization region. Calculations performed over the full surface Brillouin zone using a \(1000\times1000\) mesh further confirmed the strong localization of the Berry curvature, although the restricted integration domain provided better resolution around the hybridization gap.
Since the Dirac cone of the bottom surface remains gapless, the slab is metallic and its Hall conductivity is not quantized. 
We therefore integrate the Berry curvature of all states below a fixed energy, $\Omega_{\rm occ}(\mathbf k)$, over the surface Brillouin zone, $C=\frac{1}{2\pi}\int d^2k\,\Omega_{\rm occ}(\mathbf k)$. 
With the energy in the magnetic gap this gives $C_{\rm  mag}\simeq-\tfrac12$, and with the energy in the hybridization gap $C_{\rm hyb}\simeq1.49$. 
To locate this response in the slab, we insert in the Kubo formula the projector $\Pi_l$ onto the Wannier orbitals of quintuple layer $l$ or of the molecule\cite{PhysRevB.84.241106,PhysRevB.98.245117},  
\begin{equation}
  \Omega_l(\mathbf k)=-2\,{\rm Im}\sum_{v,v'\in{\rm occ}}\sum_{c\notin{\rm occ}}
  \frac{\langle v|\Pi_l|v'\rangle\langle v'|\partial_{k_x}H|c\rangle\langle c|\partial_{k_y}H|v\rangle}{(E_{v'}-E_c)(E_v-E_c)},
  \qquad C_l=\frac{1}{2\pi}\int d^2k\,\Omega_l(\mathbf k).
\end{equation}
Since the projectors $\Pi_l$ add up to the identity, $\sum_lC_l=C$. 
With the energy in the hybridization gap [Fig.~3(c) of the main text], the molecule and the topmost quintuple layer contribute $C_{\rm NO}\simeq0.78$ and $C_{\rm QL5}\simeq0.58$, the next layer $0.13$, and the remaining layers $|C_l|\lesssim0.06$, with the bottom layer giving $\simeq-0.05$. 
The Hall response is thus carried by  the functionalized surface, whose contribution, $\simeq1.49$, corresponds to a surface Hall conductivity $\sigma_{xy}\simeq\tfrac32\,e^2/h$. The deviation from $\tfrac32$ reflects the finite $\mathbf  k$ mesh and the small contribution of the gapless bottom surface.

\subsubsection{Molecule--cone coupling from the Wannier Hamiltonian [Eq.~(15)]}
\label{sec:wannier-phase}
We choose a single, $\mathbf k$-independent reference orbital $|{\rm NO}\rangle$ as the SOMO eigenvector of the isolated molecular block of $H_W(\bar\Gamma)$. At each point $\mathbf k$ on a loop enclosing $\bar\Gamma$, we form the projector $P(\mathbf k)=\sum_{n\in\{{\rm NO},{\rm cone}\}}|u_n\rangle\langle u_n|$ onto the two hybridizing bands. Within $160$~meV of $\varepsilon_F$, the NO-dominant band is the one with the largest weight on the N and O orbitals, and the cone band is the one, among the remaining states, with the largest weight on the top quintuple layer; the second condition excludes the bottom-surface cone, which crosses the SOMO almost without coupling to it. Writing $|{\rm cone}\rangle\equiv|u_{\rm cone}(\mathbf k)\rangle$, the orthonormal frame of Eq.~(15) of the main text is
\begin{equation}
|M\rangle=\frac{P|{\rm NO}\rangle}{\|P|{\rm NO}\rangle\|},\qquad
|D\rangle=\frac{(P-|M\rangle\langle M|)|{\rm cone}\rangle}
{\|(P-|M\rangle\langle M|)|{\rm cone}\rangle\|},
\end{equation}
and the coupling is $h_{DM}(\mathbf k)=\langle D|H_W(\mathbf k)|M\rangle$. The construction requires nonzero denominators and a continuously isolated two-band subspace. The phase of $|M\rangle$ is anchored to the fixed reference $|{\rm NO}\rangle$.

The phase of $|D\rangle$ is fixed as for $|u_+\rangle$ in Eq.~\eqref{eq:SI-dirac-spinors}, whose $|\psi_{+\frac12}\rangle$ component, $\cos(\theta/2)$, is real and positive. We take as reference $|\psi_{+\frac12}^{\rm ref}\rangle$ the $J_z=+\tfrac12$ member of the top-surface Dirac doublet of $H_W$ at $\bar\Gamma$, $\simeq211$~meV below $\varepsilon_F$: we remove its components on the N and O orbitals, orthonormalize the doublet, and diagonalize $J_z=L_z+s_z$ of the $p$ shells within it. At each $\mathbf k$ we then require $\langle\psi_{+\frac12}^{\rm ref}|D\rangle$ to be real and positive. Because this condition is imposed point by point, rather than by transport along the loop, the gauge is single-valued (periodic in $\varphi$), and the winding of $h_{DM}$ on a closed loop is an integer. It is well defined wherever $\langle\psi_{+\frac12}^{\rm ref}|D\rangle\neq0$ and $h_{DM}\neq0$. In this gauge $h_{DM}$ is the \textit{ab initio} counterpart of $g$ in Eq.~(3) of the main text.

We evaluate $h_{DM}$ on circles $|\mathbf k|=k_0$ around $\bar\Gamma$, with $k_0=0.06$ and $0.08$~\AA$^{-1}$, using $360$ points each. On both loops $|\langle\psi_{+\frac12}^{\rm ref}|D\rangle|\geq0.52$ and $|h_{DM}|\geq6$~meV, and the closed-loop sum of phase increments gives $\operatorname{wind}h_{DM}=+1.000$. Hence $n_g=+1$ and, by Eq.~(7) of the main text with $n\to n_g$, $C_{\rm hyb}=n_g+\tfrac12=\tfrac32$, in agreement with the Berry-curvature integration. 

\subsubsection{Effective couplings from the Wannier Hamiltonian}
\label{sec:SI-downfolding}
To estimate the model parameters $\alpha_\pm$ and $\beta_\pm$ from the Wannier hamiltonian $H_W$, we observe that the weights on the angular harmonics of $g$ [Eqs.~(12) and (13) of the main text, Eq.~\eqref{eq:SI-g-general}] obey
\begin{equation}
g_n(k_r)\equiv\frac{1}{2\pi}\oint d\varphi\,e^{-in\varphi}\,g(k_r,\varphi):\qquad
g_{+1}=k_r\cos\tfrac\theta2\,u\,\tilde\alpha_+,\qquad
g_{-1}=k_r\cos\tfrac\theta2\,w\,\tilde\alpha_-,\qquad
g_{n\neq\pm1}=0.
\label{eq:SI-harmonics-model}
\end{equation}
Their \textit{ab initio} counterparts are the angular harmonics of $h_{DM}$ in the periodic gauge of Sec.~\ref{sec:wannier-phase}, in which $h_{DM}^{\rm per}$ plays the role of $g$.

The values of $g_n$ can be evaluated using $h_{DM}^{\rm per}$ in place of $g$. 
The weights $u$, $w$ are those of the reference orbital $|{\rm NO}\rangle$: its projection onto the $p_{\pm1}\!\uparrow$ combinations of the in-plane N orbitals gives $|u|^2=0.49$ and $|w|^2=0.51$. 

\begin{figure}[t]
\centering
\includegraphics[width=0.95\columnwidth]{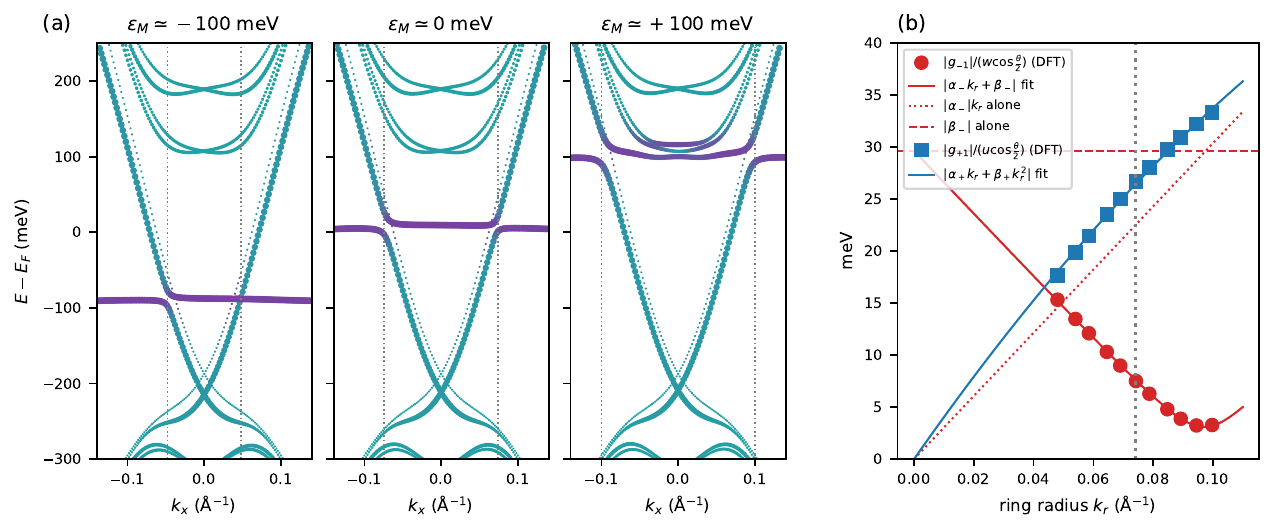}
\caption{(a) Bands of $H_W$ along $k_x$ for three rigid shifts of the in-plane NO block, placing the SOMO at $\epsilon_M\simeq-100$, $0$ and $+100$~meV. Color: relative weight on the NO orbitals (purple) and on the top quintuple layer (turquoise); symbol size: total weight on both. Dotted lines: the ring radius $\pm k_r$ used in (b). (b) Ring couplings vs.\ $k_r$ along the same scan. Red: the $m_\ell=-1$ channel (dots) with the fit $|\alpha_-k_r+\beta_-|$ (solid) and the two terms separately (dotted, dashed). Blue: the $m_\ell=+1$ channel with the fit $|\alpha_+k_r+\beta_+k_r^2|$. Vertical dotted line: the physical ring.}
\label{fig:SI-downfolding}
\end{figure}

On the ring $E\gg|m|$, so $\cos(\theta/2)\simeq1/\sqrt2$ and $\tau\simeq1$, and Eq.~\eqref{eq:SI-harmonics-model} becomes [see also Eq.~\eqref{eq:SI-g-general}]
\begin{equation}
\frac{g_{-1}}{w\cos(\theta/2)}=k_r \tilde\alpha_-=\alpha_-k_r+\beta_-,\qquad
\frac{g_{+1}}{u\cos(\theta/2)}=k_r \tilde\alpha_+=\alpha_+k_r+\beta_+k_r^2 .
\end{equation}
At a fixed ring radius $k_r$, $g_{\pm1}$ determine only the combinations $\tilde\alpha_\pm$. Since the two routes into each channel, through $|\psi_{+\frac12}\rangle$ ($\alpha_\pm$) and through $|\psi_{-\frac12}\rangle$ ($\beta_\pm$), scale with different powers of $k_r$, they can be separated by varying $k_r$ and fitting the dependence of $g_{\pm1}$ on it, provided the couplings do not depend appreciably on energy over the scanned window.  
We vary $k_r$ by shifting the on-site energies of the in-plane NO block rigidly, which leaves $u$ and $w$ unchanged. This moves the SOMO from $-100$ to $+100$~meV relative to $\varepsilon_F$, and the ring from $k_r=0.048$ to $0.100$~\AA$^{-1}$ [Fig.~\ref{fig:SI-downfolding}(a)], where $k_r$ is the radius at which the SOMO weight is split equally within the hybridizing pair. 
A complex linear least-squares fit over the scan gives 
\begin{center}
\begin{tabular}{lccc} 
\hline
channel & $|\alpha|$ (meV\,\AA) & $|\beta|$ & relative phase\\\hline
$m_\ell=-1$ & $|\alpha_-|=304$ & $|\beta_-|=29.7$~meV & $\arg(\beta_-/\alpha_-)=3.04$\\
$m_\ell=+1$ & $|\alpha_+|=411$ & $|\beta_+|=739$~meV\,\AA$^2$ & $\arg(\beta_+/\alpha_+)=-3.05$\\\hline
\end{tabular}
\end{center}
with relative residuals of 0.9\% and 0.8\% [Fig.~\ref{fig:SI-downfolding}(b)], which supports the assumption of energy-independent couplings. 
With $|u|\cos(\theta/2)\simeq|w|\cos(\theta/2)\simeq\tfrac12$, the two harmonics of $g$ are $\simeq13$ and $\simeq4$~meV, so $n_g=+1$, as found directly in Sec.~\ref{sec:wannier-phase}.

At $k_r\simeq0.074$~\AA$^{-1}$, $|\alpha_-|k_r=22.5$~meV and $|\beta_-|=29.7$~meV combine to $|\tilde\alpha_-|k_r\simeq7.7$~meV, a threefold suppression, whereas in the $m_\ell=+1$ channel $|\beta_+|k_r\simeq55$~meV\,\AA{} reduces $|\alpha_+|$ only by $\simeq13\%$, to $|\tilde\alpha_+|k_r\simeq26$~meV. 
The two routes scale differently with the ring radius: in the $m_\ell=-1$ channel the constant $\beta_-$ competes with $\alpha_-k_r$, so its relative weight grows as the ring shrinks, whereas in the $m_\ell=+1$ channel $\beta_+$ enters $\tilde\alpha_+$ as $\beta_+k_r$ and becomes negligible at small $k_r$. 
The relevance of $\beta_{\pm}$ is therefore controlled by the position of the ring, i.e.\ by the energy of the SOMO above the Dirac point, and for the present system it suppresses the $m_\ell=-1$ channel.

Notice that the extracted parameters are effective couplings of the \textit{dressed} Dirac doublet, meaning that $H_W$ also contains harmonics that $C_\infty$ forbids: $|g_0|\simeq5.6$--$7$~meV, nearly independent of $k_r$, and $|g_{\pm2}|\simeq0.1$--$0.15\,|g_{+1}|$. As the results indicate, these additional harmonics do not change $n_g$ at the physical point.

\subsubsection{Deformation of the DFT SOMO toward the chiral limit}
\label{sec:SI-deformation}
To connect the DFT SOMO with a chiral $j=\tfrac32$ orbital without closing the hybridization gap, we introduce a deformation of the Wannier Hamiltonian $H_W$ which acts only within the intramolecular on-site block.
We define
\begin{equation}
H(\lambda)=H_W-\lambda A_{\rm NO}-\lambda\delta\,L_z^{\rm NO}+s(\lambda)P_{\rm NO}^{\parallel},\qquad 0\leq\lambda\leq1 .
\label{eq:SI-deformation}
\end{equation}
Here $A_{\rm NO}$ is the part of the $\mathbf R=0$ block (within the same unit cell) among the N and O orbitals that does not conserve $J_z$, obtained by writing that block in the basis of $p_{m_\ell}$ orbitals times spin and keeping the elements between states of different $J_z$; it contains the crystal-field mixing $\Delta_{xy}$ of $p_{\pm1}$. $L_z^{\rm NO}$ is the orbital angular momentum of the N and O $p$ shells, and $\delta=0.2$~eV favors $m_\ell=+1$. $P_{\rm NO}^{\parallel}$ projects onto the in-plane ($m_\ell=\pm1$) N and O orbitals, and the shift $s(\lambda)$ keeps the SOMO of the isolated molecular block at $\bar\Gamma$ at its DFT energy, $6.7$~meV below $\varepsilon_F$, so that the inversion ring stays in place. All molecule--substrate and intermolecular hoppings are left unchanged. At $\lambda=1$ the on-site molecular block is $C_\infty$ symmetric, and the SOMO is $p_{+1}\!\uparrow$.

In Figure 3(d) of the main text, the angular harmonics $g_n$ of $h_{DM}(\lambda)$ [Eq.~\eqref{eq:SI-harmonics-model}] correspond to the loop $k_0=0.073$~\AA$^{-1}$. The hybridization gap in that figure is defined as the minimum over the annulus $0.045\leq|\mathbf k|\leq0.12$~\AA$^{-1}$ of the direct gap between the hybridizing pair of bands. Along the path the $m_\ell=+1$ harmonic dominates, $h_{DM}(\lambda)$ winds $+1$ at every step, with $|\langle\psi_{+\frac12}^{\rm ref}|D\rangle|\geq0.52$, and the hybridization gap grows monotonically from $17.4$ to $38.8$~meV. At $\lambda=1$, $|g_{+1}|\simeq23.9$~meV against $\sum_{n\neq1}|g_n|\simeq16.7$~meV, so that $|h_{DM}|\geq|g_{+1}|-\sum_{n\neq1}|g_n|\simeq7$~meV on the loop.

\bibliographystyle{apsrev4-2}
\bibliography{biblio}